\RequirePackage{fix-cm}
\documentclass{svjour3}                     
\smartqed  
\usepackage{graphicx}
\usepackage[pdftex]{color}
\newcommand{\degs}{\ifmmode ^{\circ}\else$^{\circ}$\fi}
\newcommand{\fdg}{\hbox{$.\!\!^\circ$}}
\newcommand{\amin}{\ifmmode ^{\prime}\else$^{\prime}$\fi}
\newcommand{\asec}{\ifmmode ^{\prime\prime}\else$^{\prime\prime}$\fi}
\newcommand{\farcs}{\hbox{$.\!\!^{\prime\prime}$}}  
\newbox\grsign \setbox\grsign=\hbox{$>$}
\newdimen\grdimen \grdimen=\ht\grsign
\newbox\laxbox \newbox\gaxbox
\setbox\gaxbox=\hbox{\raise.5ex\hbox{$>$}\llap
     {\lower.5ex\hbox{$\sim$}}}\ht1=\grdimen\dp1=0pt
\setbox\laxbox=\hbox{\raise.5ex\hbox{$<$}\llap
     {\lower.5ex\hbox{$\sim$}}}\ht2=\grdimen\dp2=0pt

\journalname{Experimental Astronomy}
\begin{document}

\title{GRIPS - \\
Gamma-Ray Imaging, Polarimetry and Spectroscopy\\
www.grips-mission.eu\footnote{See this Web-site for the author's affiliations.}}


\titlerunning{GRIPS}        

\author{Jochen Greiner \and
Karl Mannheim \and \\
Felix Aharonian \and
Marco Ajello \and
Lajos G. Balasz \and 
Guido Barbiellini \and 
Ronaldo Bellazzini \and 
Shawn Bishop \and
Gennady S. Bisnovatij-Kogan \and 
Steven Boggs \and
Andrej Bykov \and 
Guido DiCocco \and 
Roland Diehl \and
Dominik Els\"asser \and
Suzanne Foley \and
Claes Fransson \and
Neil Gehrels \and
Lorraine Hanlon \and
Dieter Hartmann \and
Wim Hermsen \and
Wolfgang Hillebrandt \and
Rene Hudec \and
Anatoli Iyudin \and
Jordi Jose \and
Matthias Kadler \and
Gottfried Kanbach \and
Wlodek Klamra \and
J\"urgen Kiener \and
Sylvio Klose \and
Ingo Kreykenbohm \and
Lucien M. Kuiper \and
Nikos Kylafis \and
Claudio Labanti \and
Karlheinz Langanke \and
Norbert Langer \and
Stefan Larsson \and
Bruno Leibundgut \and
Uwe Laux \and
Francesco Longo \and
Kei'ichi Maeda \and
Radoslaw Marcinkowski \and
Martino Marisaldi \and
Brian McBreen \and
Sheila McBreen \and
Attila Meszaros \and
Ken'ichi Nomoto \and 
Mark Pearce \and
Asaf Peer \and
Elena Pian \and
Nikolas Prantzos \and
Georg Raffelt \and
Olaf Reimer \and
Wolfgang Rhode \and
Felix Ryde \and
Christian Schmidt \and
Joe Silk \and
Boris M. Shustov \and
Andrew Strong \and
Nial Tanvir \and
Friedrich-Karl Thielemann \and
Omar Tibolla \and
David Tierney \and
Joachim Tr\"umper \and
Dmitry A. Varshalovich \and
J\"orn Wilms \and
Grzegorz Wrochna \and
Andrzej Zdziarski \and
Andreas Zoglauer
}

\authorrunning{Greiner \& Mannheim et al.} 

\institute{Jochen Greiner \hfill Karl Mannheim\hspace{5.05cm} \at
     MPI f\"ur extraterrestrische Physik \hfill Inst. f. Theor. Physik \& Astrophysik, Univ. W\"urzburg\\
     85740 Garching, Germany \hfill 97074 W\"urzburg, Germany\hspace{3.55cm} \\
     Tel.: +49-89-30000-3847 \hfill Tel.: +49-931-318-500\hspace*{4.28cm}\\
     \email{jcg@mpe.mpg.de \hfill E-mail: mannheim@astro.uni-wuerzburg.de\hspace{1.48cm}} 
}

\date{Received: 21 April 2011 / Accepted: 2011}

\maketitle


\begin{abstract}
We propose to perform a continuously scanning 
all-sky survey from 200 keV to 80 MeV achieving a sensitivity which is 
better by a factor of 40 or more compared
to the previous missions in this energy range (COMPTEL, INTEGRAL;  see Fig. \ref{fig_sensitivity}). 
The Gamma-Ray Imaging, Polarimetry and Spectroscopy (``{\bf GRIPS}'') 
mission addresses fundamental questions 
in ESA's Cosmic Vision plan. Among the major themes of the strategic 
plan, GRIPS has its focus on the evolving, violent Universe, exploring
a unique energy window.  We propose to investigate $\gamma$-ray bursts
and blazars, the mechanisms behind supernova explosions, 
nucleosynthesis and spallation, the enigmatic origin of positrons in 
our Galaxy, and the nature of radiation processes and particle acceleration in 
extreme cosmic sources including pulsars and magnetars.   The natural energy 
scale for these non-thermal processes is of the order of MeV.  Although they can be 
partially and indirectly studied using other methods, only the proposed 
GRIPS  measurements will provide direct access to their primary 
photons.  GRIPS will be a driver for the study of transient sources in the era
of neutrino and gravitational wave observatories such as IceCUBE and LISA, 
establishing a 
new type of diagnostics in relativistic and nuclear astrophysics.
This will support extrapolations to investigate star formation, 
galaxy evolution, and black hole formation at high redshifts.
\end{abstract}

\section{Introduction}
\label{intro}

Photon energies between hard X-rays of 0.2\,MeV and $\gamma$-rays of 
80\,MeV cover the range where many of the most-spectacular cosmic sources have their peak emissivity, so that
essential physical processes of high-energy astrophysics can be studied most directly.
Moreover, excitation and  binding energies of atomic nuclei fall 
in this regime, which therefore is as important for high-energy
astronomy as optical astronomy is for phenomena related to atomic physics.
In addition, it includes the energy scale of the electron and pion rest mass.
Current instrument sensitivities expose an ``MeV-gap'' 
exactly over this range (Fig.~1, left). The GRIPS mission will improve the 
sensitivity in this gap by a factor of 40 compared to previous missions.
Therefore, the GRIPS all-sky survey with $\gamma$-ray imaging, 
polarimetry, and spectroscopy promises new discoveries, beyond its 
precision diagnostics of primary high-energy processes.

\begin{figure}[th]
  \vspace*{-0.3cm}
    \includegraphics[width=0.42\textwidth]{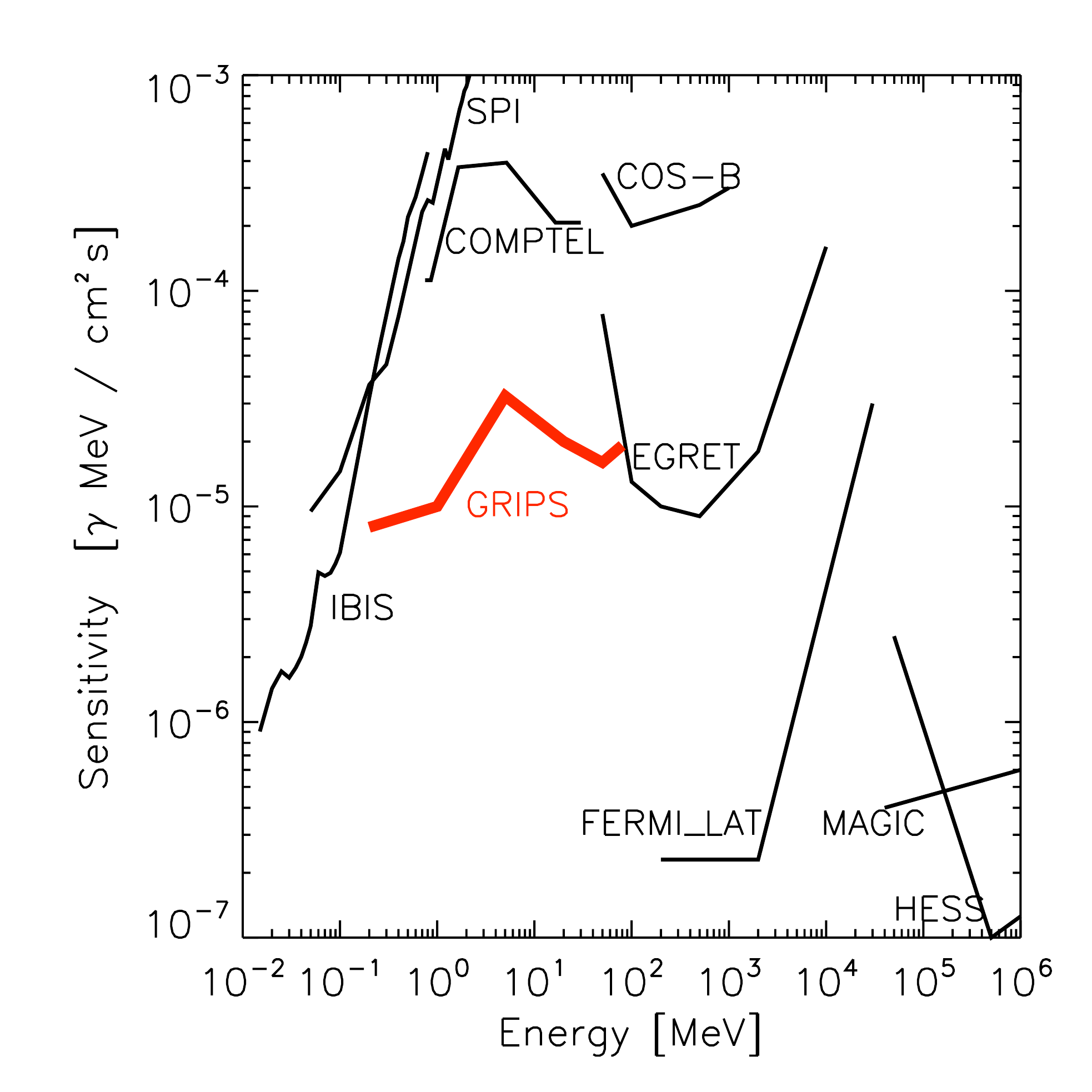}
    \includegraphics[width=0.6\textwidth]{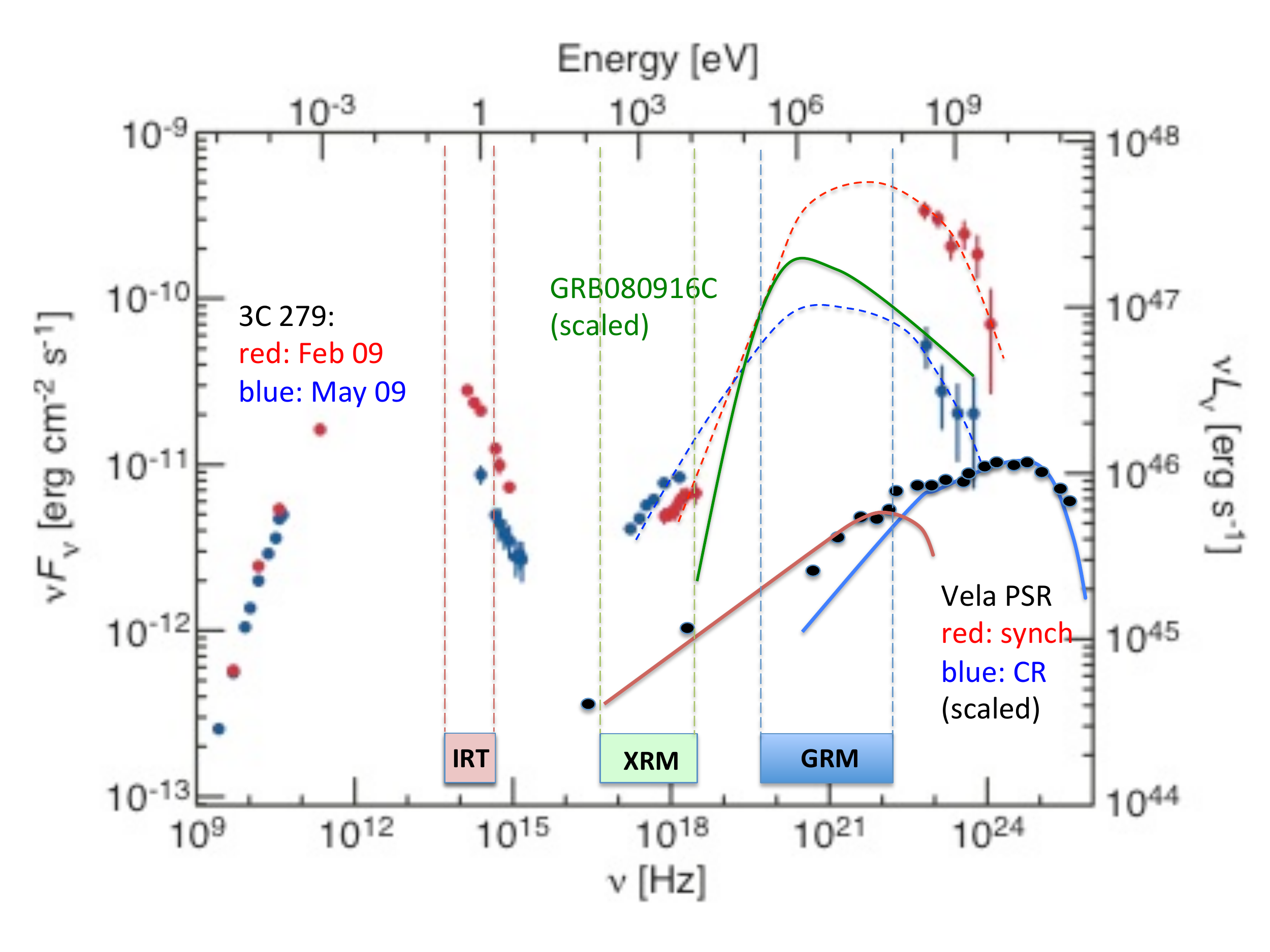}
  \caption{{\bf Left:} GRIPS will allow a major 
sensitivity improvement
in an energy range (between hard X-rays and GeV $\gamma$-rays) which 
has been poorly explored, yet holds unique information for a wide range 
of astrophysical questions. The curves are for an exposure of 10$^6$ sec,
$\Delta E = E$, and an $E^{-2}$ spectrum.
{\bf Right:} Not only GRBs, 
but also blazar SEDs peak in the MeV range,
and pulsars turn over from their maximum in the {\it Fermi} band.
The combined $\gamma$-, X-ray and near-infrared coverage of blazars
covers both emission components simultaneously. }
\vspace{-0.3cm}
\label{fig_sensitivity}
\end{figure}

GRIPS would open the astronomical window to the bizarre and highly variable 
high-energy sky, to investigate fascinating cosmic objects 
such as $\gamma$-ray bursts, blazars, supernovae and their remnants, 
accreting binaries with white dwarfs, neutron stars or black holes
often including relativistic jets,
pulsars and magnetars, and the often peculiar cosmic gas in their surroundings. 
Many of these objects show MeV-peaked spectral energy distributions 
(Fig.~1, right) or 
characteristic spectral lines; we target such primary emission to understand 
the astrophysics of these sources. 

Unrivaled by any other method, the detection of highly penetrating 
$\gamma$-rays from cosmological $\gamma$-ray bursts 
will shed  light on the first massive stars and galaxies which formed in 
obscuring gas clouds during the dark ages of the early Universe.
Polarization measurements of $\gamma$-ray bursts and blazars will for the 
first time decipher the mechanism of jet formation in
accreting high-spin black hole systems ranging from stellar to galactic masses.

The primary energy source of supernova (SN) light is radioactive decay, 
deeply embedded below the photosphere as it appears in conventional astronomical bands. 
The first direct measurement of the nickel and cobalt decay inside Type~Ia SNe 
will pin down their explosion physics and
disentangle their progenitor channels. This will impact the
luminosity calibration of Type~Ia SNe that serve as standard candles in cosmology.
Similarly, the otherwise unobtainable direct measurement of the inner ejecta and the explosive 
nucleosynthesis of core
collapse supernovae will allow to establish a 
physical model for these important terminal stages of massive-star evolution.
Explosion asymmetries \cite{2010ApJ...714.1371H} and the links to long GRBs are 
important aspects herein.
Pair-instability supernovae from very massive stars will be unambiguously 
identified through
their copious radioactivity emission.
These observations will be crucial for complementing  
neutrino and gravitational wave measurements, and for our understanding 
of the astrophysical processes and sources which underly and generate cosmic chemical evolution.

Nuclear de-excitation lines of abundant isotopes like $^{12}$C and $^{16}$O, 
the hadronic fingerprints of cosmic-ray acceleration, 
are expected to be discovered with GRIPS.
Understanding the relative importance of leptonic and hadronic processes, 
and the role of cosmic rays in heating and ionizing molecular clouds and thus 
seeding interstellar chemistry
will boost our understanding of both relativistic-particle acceleration and the 
cycle of matter.
The detection of instabilities in the supercritical magnetospheres
of magnetars, which are expected to lead to few-hundred keV to possibly MeV-peaked 
emission, 
will explore white territory on the field of plasma physics.
Resolving the riddle of the positron annihilation line will shed new light on dark 
matter annihilation 
and other sources of anti-matter in the Galaxy.

\begin{figure}[th]
\vspace{-0.2cm}
\includegraphics[width=0.5\columnwidth]{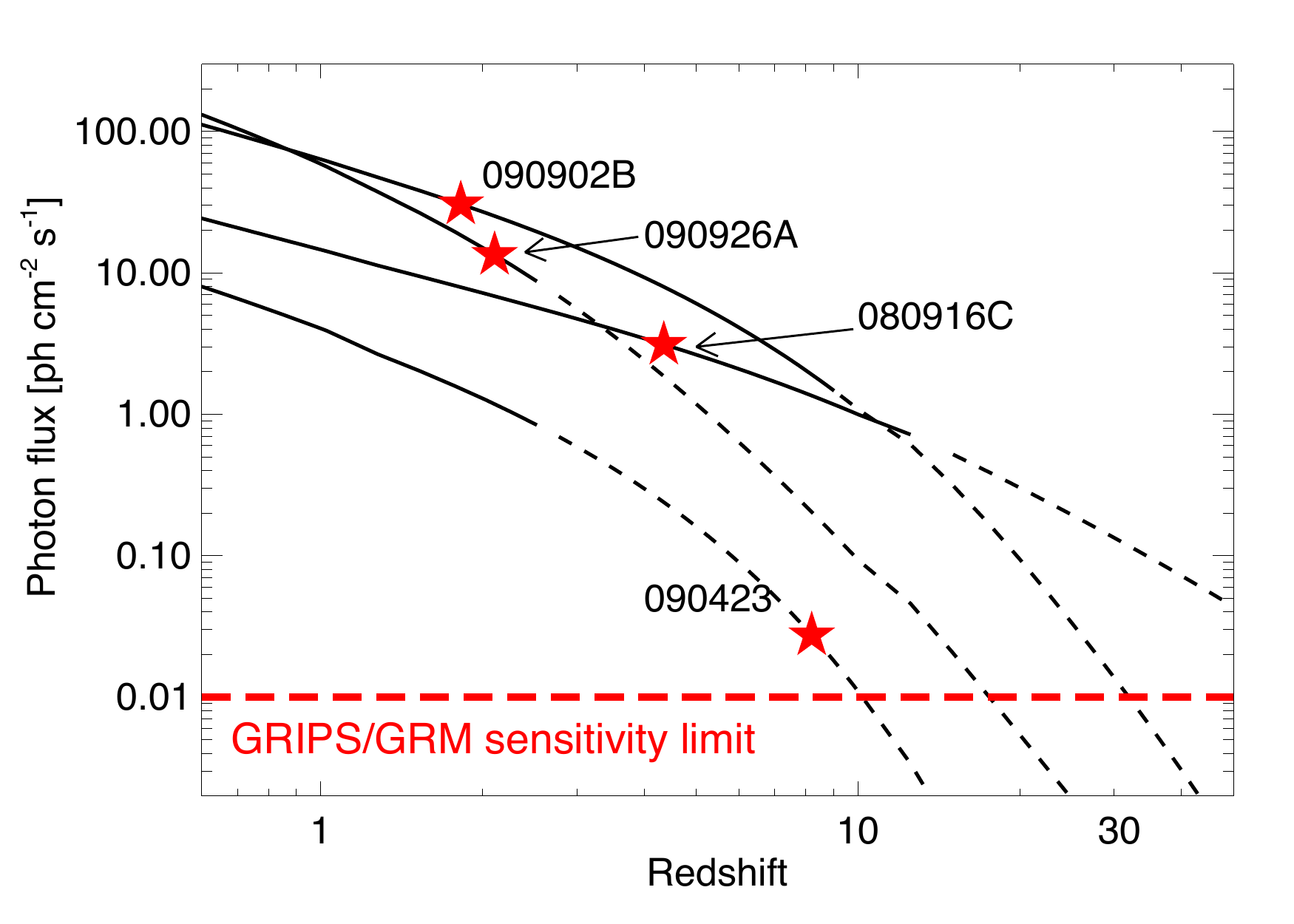}
\hfill\parbox[t]{6.cm}{\vspace*{-4.3cm}\caption{The 200\,keV--2\,MeV GRB 
peak photon flux as a function of 
burst redshift. The positions of three {\it Fermi}/GBM+LAT GRBs as 
well as the highest redshift GRB\,090423 are indicated (stars) with 
their predicted peak fluxes. The crossing of the dashed curves with 
the sensitivity curve (horizontal red line; 1\,s,  30\degs\ off-axis)
 marks the largest redshift
up to which these bursts would have been detectable with GRIPS.
The transition between solid and dashed curves marks the redshift
below which also $E_{\rm peak}$ can be measured with GRIPS.}
\label{flux_z}}
\vspace{-0.8cm}
\end{figure}

We detail in the following sub-sections how GRIPS will answer the burning questions:
{\it How do stars explode?}
{\it What is the structure of massive-star interiors and of compact stars?}
{\it How are cosmic isotopes created and distributed?}
{\it How does cosmic-ray acceleration work?}
{\it How is accretion linked with jets?}
Answering these questions will provide the basis to understand the larger 
astrophysical scales,
like the interstellar medium as it evolves in galaxies, the supernova-fed intergalactic gas in 
galaxy clusters, 
and the cosmic evolution of elemental abundances.

\section{Scientific Objectives}

\subsection{Gamma-Ray Bursts and First Stars}

GRIPS will observe in the energy range where GRB emission peaks.
With its energy coverage up to 80 MeV, GRIPS will firmly establish 
the high energy component seen in {\it CGRO}/EGRET ($>$10 MeV) and  
{\it Fermi}/LAT bursts ($>$100\,MeV) in  much  larger numbers,  
and characterize its origin through polarisation signatures. 
GRIPS will measure the degree of polarisation of the
prompt $\gamma$-ray burst emission to a few percent accuracy for more
than 10\% of the detected GRBs (see simulations in \S~4), and securely
measure how the degree of polarisation varies with energy
and/or time over the full burst duration for dozens of bright GRBs.
Also, the delay of GeV photons relative to emission at $\sim$ hundred keV,
observed in a few GRBs with {\it Fermi}/LAT, manifests itself already
at MeV energies in {\it Fermi}/GBM, and will thus be a science target
for GRIPS. These observations enable a clear identification of the prompt 
GRB emission processes, and determine the role played by magnetic fields.

Due to its outstanding sensitivity (Fig.~2), GRIPS will detect 
$\sim$ 650 GRBs yr$^{-1}$
and measure the incidence of gas and metals through X-ray absorption
spectroscopy and line-of-sight properties by enabling NIR spectroscopy
with {\it JWST}. GRIPS should detect more than 20 GRBs at $z > 10$
over 10 years, and the IRT will determine  photometric redshifts for  
the bulk  of the $z>7$ sources (Fig.~3).

\begin{figure}[h]
\vspace{-.2cm}
\centering{
\includegraphics[width=0.7\textwidth]{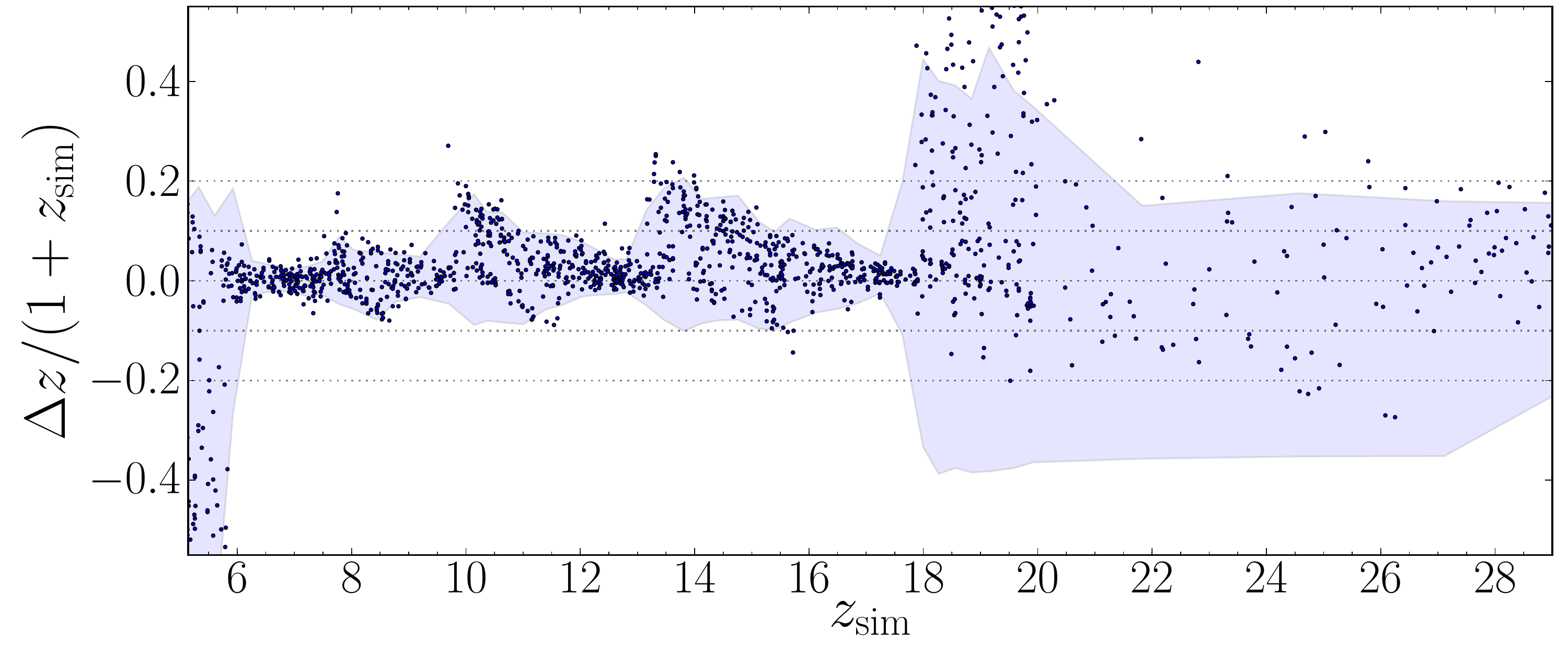}}
\vspace{-0.2cm}
\caption{GRB afterglow photometric redshift accuracy of the IRT filter set 
as in Fig. \ref{filters}. Black dots show 900 simulated afterglows and
shaded area represents the 1$\sigma$ statistical 
uncertainty of the photo-$z$ analysis averaged over 30 afterglows in 
relative ($\eta = \Delta z/(1+z)$) terms. 
For the $7<z<17$ redshift range, the photo-z can be determined to better
than 20\%. At $z > 17.5$ (K-dropout), the error gets larger due to
the gap above the $K$ band and the widths of the $L$ ($M$) bands;
yet, the redshift accuracy is more than sufficient for any follow-up decision.}
\label{IRTaccuracy}
\end{figure}

If  the  GRB  environments contain total hydrogen column densities of 
10$^{25}$\,cm$^{-2}$, or higher, GRIPS holds the promise of measuring  
redshifts directly from the $\gamma$-ray spectrum via nuclear resonances, 
and will be sensitive to do so beyond z$\sim$13.

GRIPS will also detect a handful of short GRBs at $z < 0.1$, enabling a
potential discovery of correlated gravitational-wave and/or neutrino signal.

\subsection{Blazars}

GRIPS will catalogue about 2000 blazars, thus probing blazar evolution to large
redshifts. These observations will pinpoint the most massive halos at large
redshifts, thus severely constraining models of structure evolution.
This large sample of blazars will establish their (evolving) luminosity
function and thus determine the fractional contribution of blazars to the
diffuse extragalactic background. GRIPS is expected to detect $\sim$10
blazars at z $>$ 8.

Studies of their nonthermal radiation mechanisms will be supported
through spectro-polarimetric measurements. The link between the inner 
accretion disk and the jet can be probed with correlated variability 
from the thermal to the nonthermal regime, using GRIPS auxiliary 
instruments. This will localize the dominant region of high-energy emission.

Through nuclear lines detected in nearby AGNs, and through tracing 
variability, GRIPS will probe the injection of accelerated particles 
into the jet plasma.

\begin{figure}[ht]
\includegraphics[width=0.55\columnwidth]{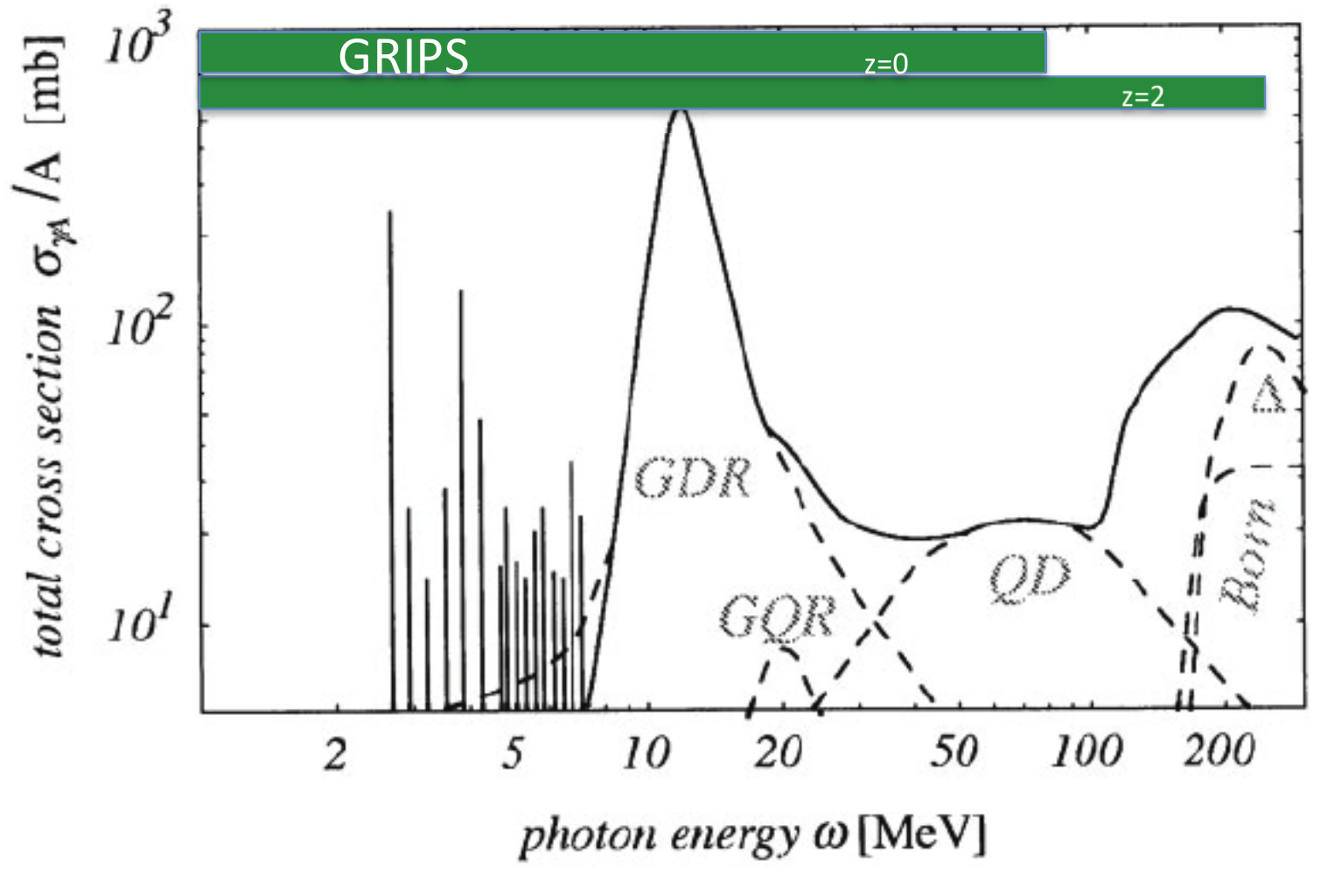}
\vspace{-0.3cm}
\hfill\parbox[t]{5.2cm}{\vspace*{-4.5cm}\caption{Nuclear interactions with photons occur in the range of nuclear
excitation levels ($\sim$MeV), collective-nucleon resonances such as Pygmy
(5--10 MeV) and Giant Dipole resonances (15-30 MeV), and individual
nucleon excitations at higher energies such as the Delta resonance.
GRIPS energies cover these line features, and thus
provide a capability for measurements of fully-ionized matter in extreme
plasma, when atomic transitions are absent.}}
\label{fig_454_spec} 
\end{figure}

\subsection{Supernovae and Nucleosynthesis}

GRIPS will search the nearby Universe out to 20~Mpc for $^{56}$Ni decay
$\gamma$-ray lines from SNIa, with an expectation of 10--20 significant detections.
Establishing ratios of various lines from the $^{56}$Ni decay chain (see
Fig.~5), and their variation with time, are key GRIPS 
objectives. Even if the lines are significantly Doppler-broadened, 
the 0.1--3~MeV continuum can be used to test different explosion scenarios 
(e.g. \cite{2008MNRAS.385.1681S}).
Combined with optical-IR data, one can determine/constrain unburnt WD material
\cite{2003ApJ...591..316M}, and with annihilation lines from $^{56}$Co and
$^{48}$V decay positrons, one has a sensitive probe (via e$^+$ propagation) of
the magnetic field structure (combed vs. scrambled) in expanding SNIa remnants.

GRIPS will also detect several (3--5) nearby core-collapse supernovae (ccSNe).
As for SNIa and SN1987A, the $\gamma$-ray escape from the ejecta reflects 
hydrodynamic large scale mixing during and after the explosion. Comparing 
$\gamma$-ray characteristics of
different classes of SNe, including pair-instability SN (due to high
($\sim$3~M$_{\odot}$) $^{56}$Ni mass \cite{2009Natur.462..624G}), and 
hyper-energetic SNe linked to GRBs (1--2), GRIPS will probe the range of
variations, thus  offering a direct view 
of the central supernova engines, and help revealing their progenitors.

The Galactic census of ccSNe and their associated $^{44}$Ti emission, deepened
by an order of magnitude in flux by GRIPS, will sample ccSNe that occurred anywhere 
in the Galaxy during the past several centuries  (estimated rate 1/50y).
Presently, Cas A (age of 340~y) is still the only clearly-detected $^{44}$Ti source, 
while we expect to see about a handful of such young remnants. 
The (currently indirectly inferred) amount of $^{44}$Ti 
in SN1987A will be measured with GRIPS at $>$5$\sigma$ significance.
Together, this will clarify how typical or anomalous $^{44}$Ti ejection 
is in ccSNe, with direct implications for non-sphericities of core-collapses as GRBs present it in its extremes.

Novae will be GRIPS-tested for $^7$Be and $^{22}$Na emission to distances of 
$\sim$5~kpc, resulting in valuable constraints for theoretical models. 
Predicted annihilation line flashes (from the decay of $^{18}$F and $^{13}$N)
are detectable throughout the Galaxy, and since they occur prior to the
optical emission, 
enable the earliest possible detections of novae and early follow-up. 
The discovery of an unexpected high-energy emission in the symbiotic binary
V407 Cygni, attributed to shock acceleration in the ejected nova shell after 
interaction with the wind of the red giant \cite{Abdo2010Sci...329}, 
opens up new  possibilities that will be probed in more detail with GRIPS.

\begin{figure}[bh] 
\includegraphics[width=0.7\textwidth]{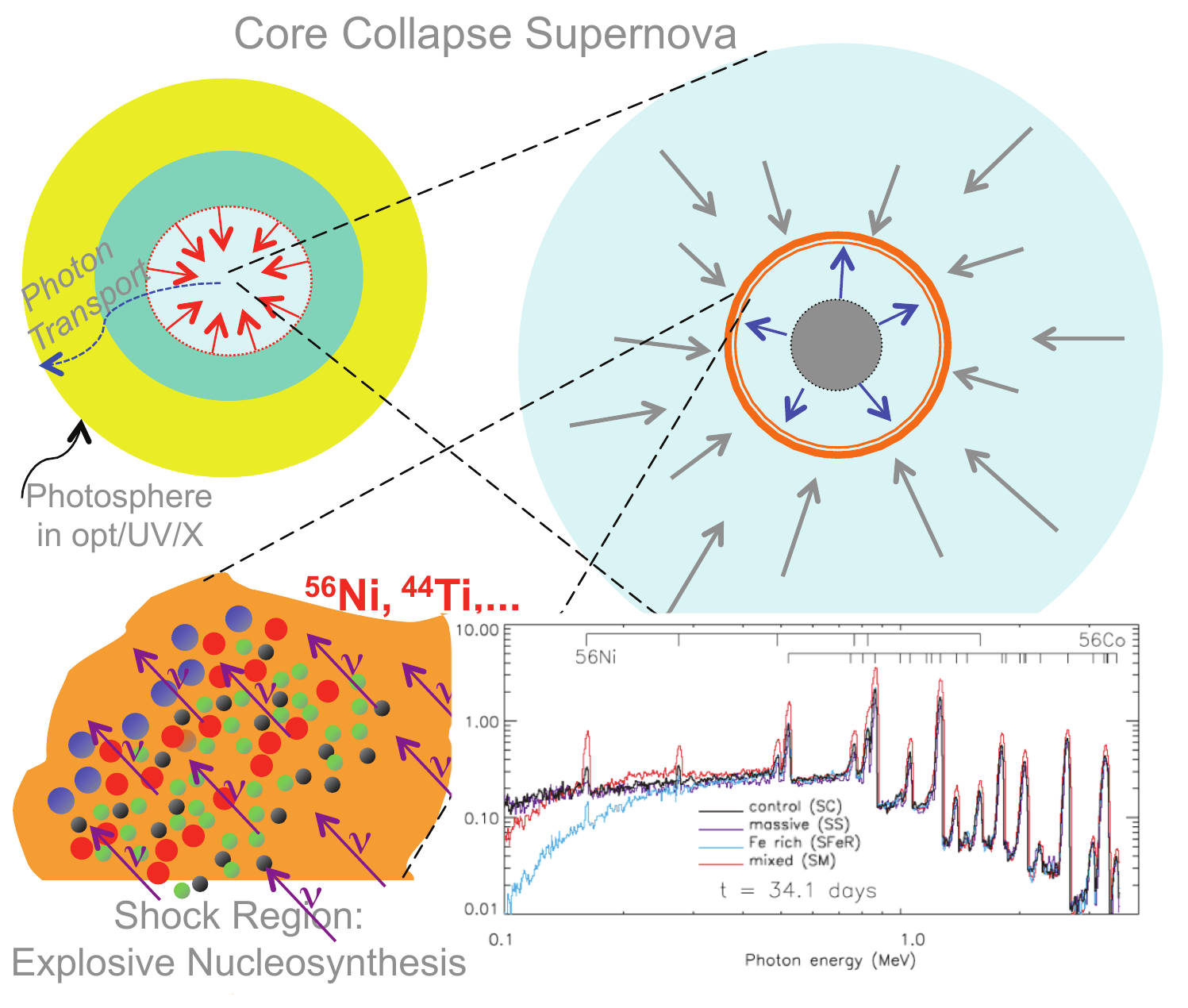}
\hfill\parbox[t]{3.3cm}{\vspace*{-7.5cm}\caption{Supernova light originates 
from $^{56}$Ni radioactivity produced in 
their inner regions. Gamma-rays from radioactivity penetrate the SN
envelope. Here a core-collapse SN is shown, with gamma-ray
emitting isotopes such as $^{56}$Ni being produced from infalling nuclei as
these are decomposed and re-assembled. Gamma-ray lines and their
changing relative intensities are direct messengers of the explosion
physics, characteristic for the different SNIa and also core-collapse
supernova variants.}}
\label{fig_SN-spec}
\vspace{-0.3cm}
\end{figure}  

\begin{figure} 
\includegraphics[width=0.6\columnwidth]{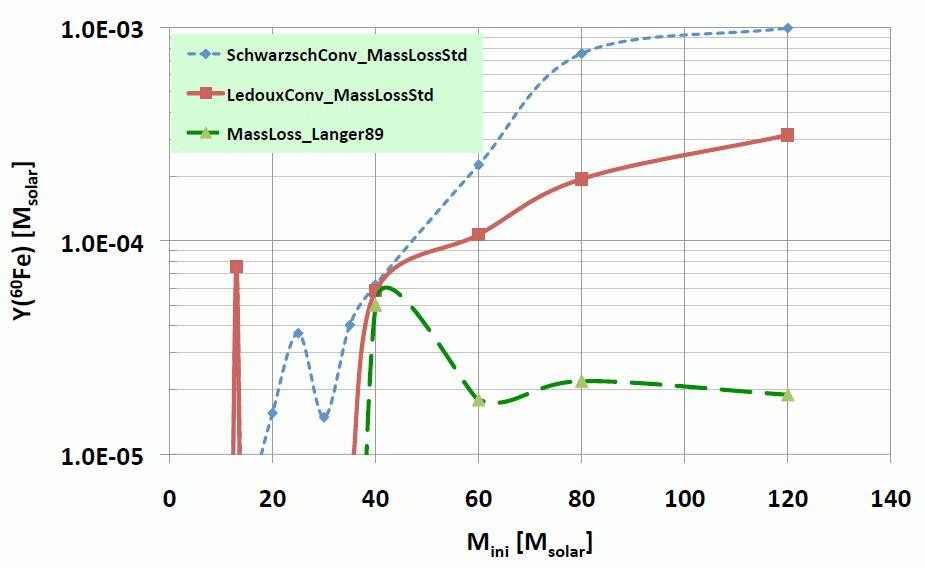}
\vspace{-0.3cm}
\hfill\parbox[t]{4.6cm}{\vspace*{-4.cm}\caption{Convection inside massive 
stars is complex, and the yields of $^{60}$Fe are sensitive to detail. 
Variations with different prescriptions of convection zones 
(Ledoux versus Schwarzschild criterion), and mass loss (reduced versus 
Langer1989), affect $^{60}$Fe yields for high-mass stars by up to an order 
of magnitude, respectively \cite{2006ApJ...647..483L}.}
\label{fig_Fe_yields}}
\end{figure}  

GRIPS will measure diffuse radioactivity afterglows of nuclear burning inside 
massive stars and supernova, at the Myr time scale ($^{26}$Al and $^{60}$Fe decays), for
dozens of nearby ($\sim$kpc) stellar associations and groups. Combined with the
stellar census for these groups as obtained from other observations (e.g. 
{\it GAIA}),
this provides unique measurements of isotopic yields, serving as calibrators
for nucleosynthesis in massive stars. These radioactivities also trace the
flow and dynamical state of hot interstellar cavities, as they merge with
the ambient ISM, their radioactive clock directly relates to the age of the
stellar group, and hence to the stellar-mass range that could have exploded
as SN. For nearby ($\leq$~1~kpc) stellar groups, the distribution 
of ejecta as compared to the parental molecular-cloud morphology thus 
provides a diagnostoc
for molecular-cloud destruction via feedback from massive stars and supernovae. GRIPS will provide a
Galactic map of $^{60}$Fe emission, essential for disentangling the 
contributions of different candidate source types to $^{26}$Al. Line ratios for specific groups will
constrain $^{60}$Fe production of massive stars beyond $\sim$40~M$_{\odot}$, 
which directly relates to (uncertain) convective-layer evolution and stellar 
rotation \cite{2006ApJ...647..483L} 
(Fig.~6).
GRIPS will also detect these two key 
radioactivities, for the first time, in the local group (M31, LMC, SMC).

\vspace{-0.3cm}

\subsection{Annihilation of Positrons}

GRIPS will probe positron escape from candidate sources along the Galactic plane
through annihilation $\gamma$-rays in their vicinity. For several microquasars
and pulsars, point-source like appearance is expected if the local annihilation
fraction $f_{local}$ exceeds 10\% (I$_{\gamma}\sim10^{-2}\cdot f_{local}$
ph~cm$^{-2}$s$^{-1}$).
GRIPS will enable the cross-correlation of annihilation $\gamma$-ray images with
candidate source distributions, such as $^{26}$Al and Galactic diffuse MeV
emission (where it is dominated by cosmic-ray interactions
with the ISM) [both also measured with GRIPS at superior quality], point sources
derived from {\it INTEGRAL}, {\it Swift}, {\it Fermi}, and {\it H.E.S.S.} 
measurements of pulsars and accreting binaries, and with candidate dark-matter 
related emission profiles.

\begin{figure}[th] 
\vspace{-0.2cm}
\includegraphics[width=0.6\textwidth]{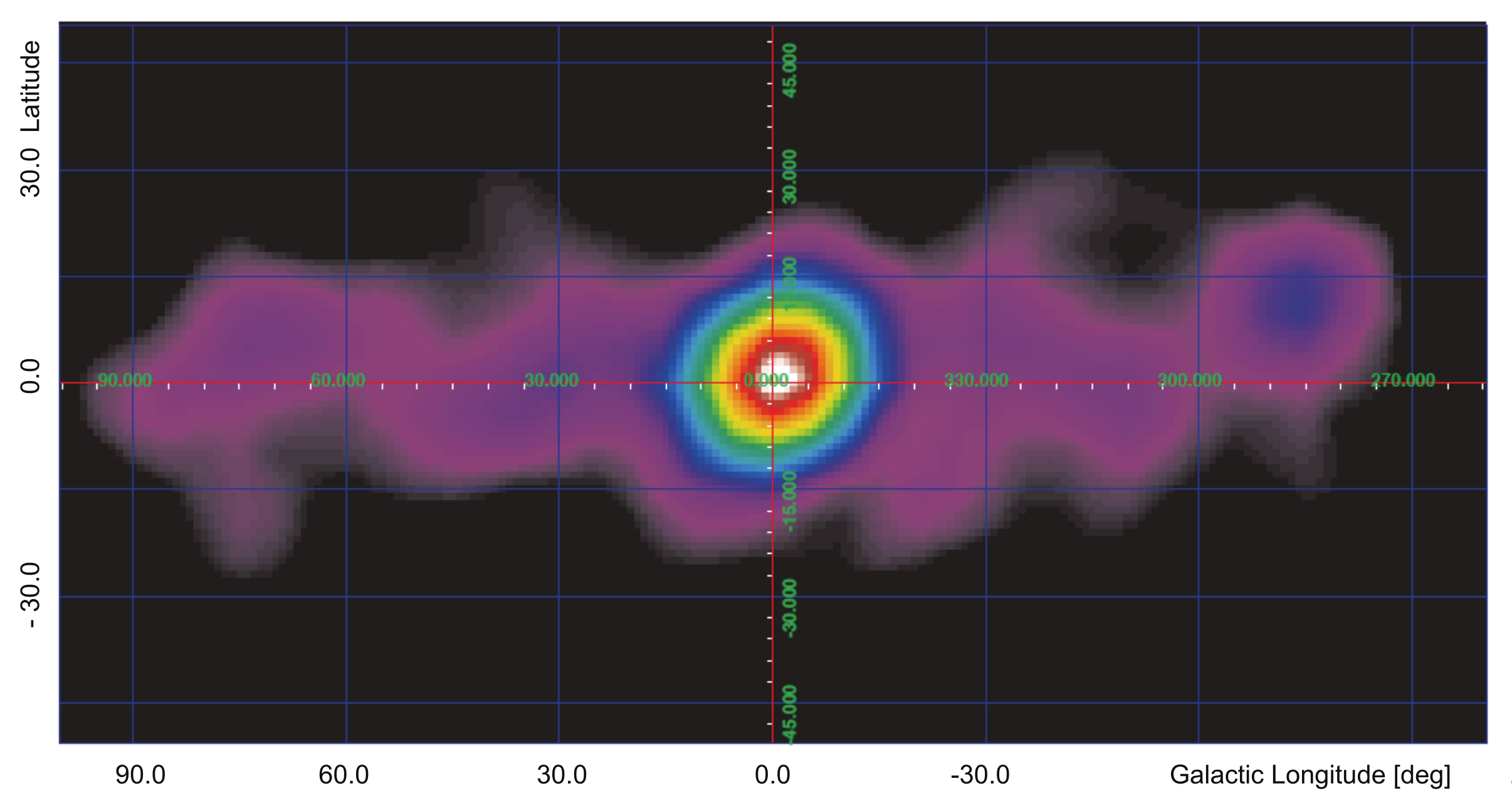}
\hfill\parbox[t]{4.2cm}{\vspace*{-2.2cm}\caption{{\it INTEGRAL/SPI} image of the annihilation emission at 511~keV 
along the inner Galaxy \cite{2010ApJ...720.1772B}, illustrating the bright, 
dominating bulge-like emission. The color scales with S/N.}}
\label{fig_511SPI}
\end{figure}  

GRIPS will deepen the current {\it INTEGRAL} sky image by at least an 
order of
magnitude in flux, at similar angular resolution. Comparing Galactic-disk and
-bulge emission, limits on dark-matter produced annihilation emission will 
constrain decay
channels from neutralino annihilation in the gravitational field of our Galaxy.
GRIPS will also be able to search for $\gamma$-ray signatures of dark
matter for nearby dwarf galaxies.

\subsection{Cosmic Rays}

GRIPS will observe diffuse nuclear de-excitation lines and 
inverse-Compton emission from the inner Galaxy. Collisions of particles with energies above 
the thermal regime are responsible for such excitation and emission, so that the acceleration
process from the thermal pool to relativistic energies and the Cosmic-Ray-ISM 
connection will be probed with unprecedented sensitivity. Furthermore, characteristic nuclear 
de-excitation lines expected from Wolf-Rayet enriched supernova remnant environments, 
could be discovered, testing one of the current CR origin models. Nuclear de-excitation
lines produced in the environments of SNRs and accreting binaries 
offer unique laboratories for gauging models of cosmic ray production, 
acceleration, transport, and interaction with their surroundings. 
GRIPS will establish this tool for particle acceleration sites in the 
Galaxy, and will also contribute to this topic in the context of 
solar flares (see section 2.7).

\subsection{Compact Stellar Objects: Pulsars, Magnetars, Accreting Binaries}

GRIPS will detect 30--40 pulsars in the 0.2--80~MeV energy range.
This estimate is based on the Fermi pulsar catalogue extrapolated to lower energies,
as well as extrapolation to higher energies of the hard spectra of
the youngest pulsars detected by {\it INTEGRAL} and {\it RXTE}/HEXTE at 
hard X-ray energies. Most of the latter pulsars appear to reach their maximum
luminosity in the MeV regime.

\begin{figure}[th]
\includegraphics[width=0.5\columnwidth]{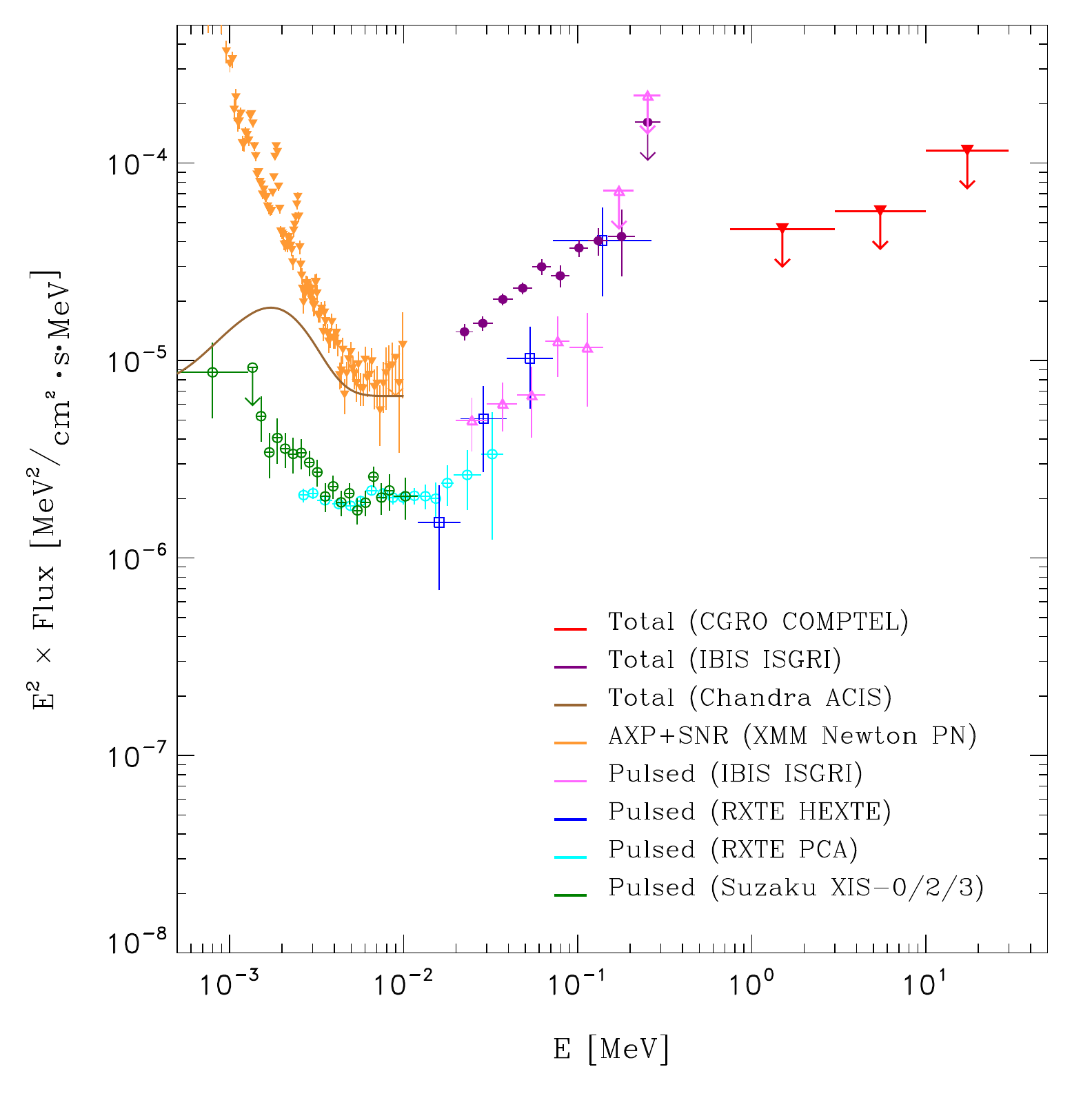}
\vspace{-0.3cm}
\hfill\parbox[t]{5.5cm}{\vspace*{-3.cm}\caption{Observed spectral energy 
distribution of SNR Kes 73 and 
its compact object AXP 1E1841-045 (total and pulsed emissions) 
This illustrates 
the importance of sensitive measurements between 100~keV and 100~MeV to
sample the break energy region.}
\label{fig_AXPs}}
\end{figure}

GRIPS-determined light-curves and phase-resolved spectra in the 0.2--80 MeV
range will provide decisive constraints on the HE-emission geometry in
pulsar magnetosphere and the acceleration processes located therein. For
brighter pulsars, polarisation data will identify the nature of the emission.

Measurements of pulsar wind nebulae (prominent examples at MeV energies are
the Crab nebula and Vela~X-1) will track the outflow of relativistic particles
and fields.

GRIPS will search for photon splitting in strong B-fields in selected pulsars
and AXPs/SGRs. Splitting suppresses the creation of pairs, and inhibits escape
of any near-surface emission above about 10--20 MeV.

GRIPS will search for long orbit neutron star/massive star binaries and for
tight accreting systems harboring neutron stars or black holes. These systems
(e.g. Cyg X-1) exhibit hard power-law spectral tails beyond 100 keV when they
are in their low/hard emission state. It is important to determine how these
hard power laws continue into the MeV range, because spectral features (lines,
breaks, cut-offs) are unique fingerprints of non-thermal particle populations
expected in those sources. GRIPS will probe accretion physics in the high-energy
domain of stellar black holes, thus aid our understanding of the more-extreme
circumstances encountered in accretion onto supermassive black holes.

\subsection{Solar Flares}

Solar flare measurements will be a natural by-product of the continuous sky survey carried
out by GRIPS as the Sun passes regularly through its field of view. 
The number
of observed flares depends on the phase of the solar cycle during the GRIPS mission.
Lightcurves in characteristic (nuclear-line and continuum) energy bands and high statistics spectra will
be obtained. Comparison of these with data from the fleet of
solar-terrestrial satellites and ground level observatories will bear on the
still somewhat limited understanding of the enigmatic flare processes and their implications for cosmic particle acceleration. 
Gamma-rays
in the MeV regime provide the means to directly study particle acceleration and
matter interactions in these magnetised, non-thermal plasmas. Polarisation
measurements are of great value for disentangling these dynamic processes.

Beyond this astrophysical reason, observing solar flares and 'space weather' are of great importance for our 
natural and technological environment. GRIPS would add an MeV-monitor to 
this program.

\newpage

\section{Mission profile}

The GRIPS mission requires that a mass of 5.1\,t be delivered into 
an equatorial, circular, low-Earth orbit (LEO) with an altitude of 500\,km.  
The primary launcher requirements are determined by the low inclination 
LEO orbit  and by the payload mass. The Soyuz Fregat 2B has a capacity of 
5.3\,t and is the only viable launch option in the Call. The GRIPS mission 
comprises two satellites containing the science instruments:  
Gamma-Ray Monitor and the X-ray/Infrared telescopes respectively. 
These can readily be accommodated within the fairing specifications
 with a diameter of 3.8 m and a height of 7.7\,m.

The orbit was selected to fulfill the following requirements in an optimized
way:
(i) Low background,
(ii) High science data fraction per orbit,
(iii) High downlink rate for data transmission
  and good longitudinal coverage of the orbit by ground stations,
(iv) High probability of the mission to remain at the chosen
orbit for a mission life-time of $\approx$10 years.
The requirement for a low background means low inclination,
preferably 0\degs\, to avoid the radiation
belts and the South Atlantic Anomaly. It is now well established
that the background in hard X-ray and $\gamma$-ray instruments in a
LEO can be a factor of 100 lower than in a Highly Eccentric Orbit,
such as that of INTEGRAL.

Both satellites of GRIPS (with an inter-satellite distance of 500--2000 km) 
should be 3-axis stabilized with closed loop 
attitude control,
following the tradition of many recent astronomical space missions,
and presenting no new problems in the control of the dynamics of the
spacecrafts. GRIPS will have, however, three distinct features:
\begin{enumerate}
\vspace{-0.2cm}\item GRIPS/GRM will do zenith-pointing all the time.
\item For each localised GRB, GRIPS/XRM\&IRT 
 should be alerted via inter-satellite link and autonomously
  slew with the co-aligned X-ray and Infrared telescopes
  towards the GRB location, similar to Swift \cite{geh04}.
  With $\approx$2 GRBs/day, and autonomous XRM\&IRT observations 
  for at least the next 24 hrs, a large fraction of the GRIPS/XRM\&IRT
  satellite operation will be to follow-up GRBs.
\item After the first $\sim$100 s X-ray observation 
  the X-ray afterglow position will be determined with an accuracy 
  of $<$45\asec, and another satellite slew commanded to place the 
  afterglow in the IRT FOV. Despite the large X-ray FOV we expect 
  no confusion of the afterglow with persistent X-ray sources for 
  two reasons: (i) Based on the X-ray afterglow intensity distribution
  at 30 min after the burst, 92\% of all known X-ray sources
  are fainter than the faintest GRB afterglow, resulting in a chance
  coincidence of 10\% to have one unrelated X-ray source in the total
  7 square-degree FOV which is brighter than the faintest X-ray 
  afterglow. (ii) From the $\sim$ 3\,yr {\it eROSITA} survey after 2014
  there will be a complete catalog of all faint X-ray sources
  (modulo variability) taken with exactly the same telescope/detector
  characteristics.
\end{enumerate}

\begin{figure}[th]
\includegraphics[width=0.45\columnwidth]{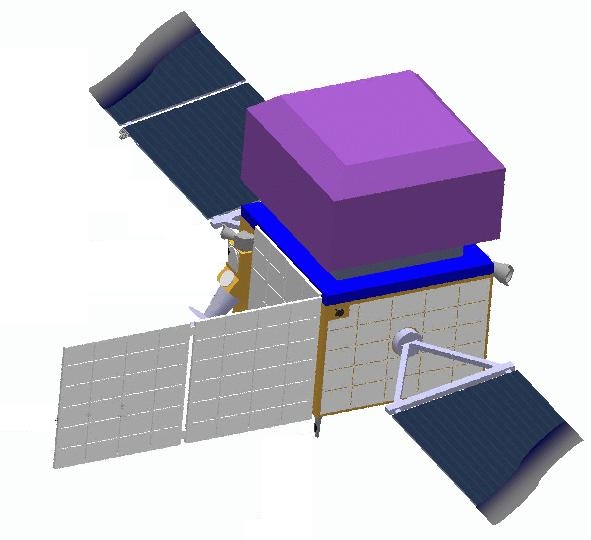}
\hfill
\includegraphics[width=0.45\columnwidth]{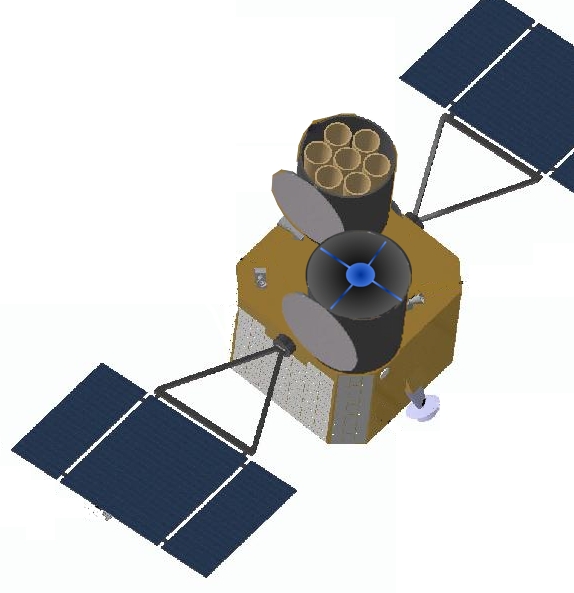}
\caption[GRIPS_2-sat]{GRIPS configuration in the two-satellite option, 
where the GRM is on one satellite (left), and XRM and IRT on the other
(right). The GRM satellite would just do the zenith scanning all-sky survey,
while XRM/IRT would re-point with the whole (second) satellite to GRBs,
similar to {\it Swift}. }
\label{fig_2sat}
\end{figure}

Highest scientific quality will be reached if all detected
photons of GRM can be telemetered to Earth with all their interaction
parameters. In GRM we expect about 68\,000 cts/s, 
thus about 1.5 Mbps. The XRM produces about 110 cts/s,
or 10.6 kbps, and the IRT about 1.2 Mbps.
Since this has to be transmitted to ground in about
8 min passes out of 96 min orbital period, we require a downlink
rate of 35 Mbps.

\section{Proposed Model Payload}

GRIPS will carry three major telescopes: the Gamma-Ray Monitor (GRM),
the X-Ray Monitor (XRM) and the Infrared Telescope (IRT). 
The {\bf GRM} is a combined Compton
scattering and Pair creation telescope sensitive in the energy range
0.2--80 MeV. It will follow the successful concepts of
imaging high-energy photons used in {\it COMPTEL} (0.7--30 MeV) and
{\it EGRET} ($>$30 MeV) but combines them into one instrument. New
detector technology and a design that is highly focused on the
rejection of instrumental background will provide greatly improved
capabilities. Over an extended energy range the sensitivity will
be improved by at least an order of magnitude with respect to
previous missions.  
In combination with improved sensitivity, the large field of view (FOV), 
better angular and
spectral resolution of GRM will allow the scientific goals
outlined in $\S$ 2 to be accomplished. 
The {\bf XRM} is based on the
mature concept and components of the {\it eROSITA} X-ray telescope,
which is scheduled for a space mission on the Russian platform
{\it Spektrum-XG} in 2013. 
The {\bf IRT} is based on the telescope as proposed for the {\it EUCLID}
mission. It uses the same main mirror and telescope structure (most
importantly distance between M1 and M2).
Instead of the suite of {\it EUCLID} instrumentation, just
one instrument is foreseen: a 7-channel imager to determine photometric
redshifts of GRB afterglows in the range $7<z<35$.

\begin{table}[hb]
\caption[Science-to-payload mapping]{Scientific requirement vs. 
payload property}
\begin{tabular}{ll}
  \hline
  \noalign{\smallskip}
large number of GRBs       & large FOV $\gamma$-ray detector \\
detect spectral lines & 3\% spectral resolution;
                             high continuum sensitivity$\!\!$ \\
detect polarisation        & record Compton events 
                               w. large scatter angle \\
onboard localisation  & computing resource \\
arcmin localisaton          & X-ray telescope \\
rapidly recognize high-z & IR telescope \\
  \noalign{\smallskip}
  \hline
  \end{tabular}
\vspace{-0.3cm}
\end{table}

\subsection{Gamma-Ray Monitor}

Two physical processes dominate the interaction of photons with
matter in the $\gamma$-ray energy  band from 200\,keV to
$\sim$80\,MeV: Compton scattering at low energies, and
electron-positron pair production at high energies, with the
crossover at $\sim$ 8\,MeV for most detector materials. In both
cases the primary interaction produces long-range secondaries
whose directions and energies must be determined to
reconstruct the incident photon properties. 

\subsubsection{Conceptual Design and Key characteristics}

The GRM will employ two separate detectors 
as was the case for previous Compton 
and pair creation telescopes. 
 The first detector is a tracker (D1), in which the initial 
Compton scatter or pair conversion takes place, and 
the other a calorimeter 
(D2), which absorbs and measures the energy of the secondaries 
(Tab. \ref{KeyChar}). 
In the case of Compton interactions, the incident photon scatters off 
an electron in the tracker. The interaction position and the 
energy imparted to the electron are measured. The scattered photon 
interaction point and energy are recorded in the calorimeter. From 
the positions and energies of the two interactions the incident 
photon angle is computed from the Compton equation. The 
primary-photon incident direction is then constrained to an event 
circle on the sky. For incident energies above about 2\,MeV the 
recoil electron usually receives enough energy to penetrate 
several layers, allowing it to be tracked. This further constrains 
the incident direction of the photon to a short arc on the event 
circle. GRM will determine GRB locations to better than 1\degs\ 
(radius, 3$\sigma$).

\begin{table}[th]
\caption[Gamma-Ray Monitor Key characteristics]{Gamma-ray Monitor 
Key characteristics \label{KeyChar}}
\begin{tabular}{clr}
  \hline
  \noalign{\smallskip}
 Detectors: &           &   $\!\!$mass+margin [kg]$\!\!$      \\
  D1        & Si DSSD   &   50+2.5    \\
            & Structure &   20+4 \\
  D2        & LaBr$_3$   &  500+25    \\
            & Structure &  240+48   \\
  ACS       & Plastic   &  130+6.5   \\
            & Structure &   40+8   \\
 Electronics &          &  420+84  \\
 \hline
  \noalign{\smallskip}
Total GRM   &           & 1578    \\
\hline
  \noalign{\smallskip}
 Channels   & D1      & 196\,608 \\
            & D2      & 104\,960 \\
            & ACS     &        8  \\
 Total      &               & 301\,576    \\
 \hline
  \noalign{\smallskip}
 Energy resolution @ 662 keV   & D1/D2  &  1.0 keV / 12.3 keV    \\
    \hline
  \noalign{\smallskip}
 Trigger thresholds @ 662 keV      & D1/D2  &  10  keV / 20 keV     \\
 \hline
 \noalign{\smallskip}
 Noise thresholds @ 662 keV     & D1/D2       & 5  keV / 10 keV        \\
  \hline
  \noalign{\smallskip}
 $\!\!$Background trigger  rate ($>$5 keV)   & LEO, i=$0^\circ$ & 68\,000 cts/s \\
  \noalign{\smallskip}
  \hline
  \noalign{\smallskip}
  ARM (FWHM) &  at 1 MeV   & 1.8 deg \\
  Localisation  &   onboard   &  $\sim$0.5 deg \\
  Cont. sensitivity @ 1 MeV & 10$^6$ s, scan   &  3$\times$10$^{-5}$ ph/cm$^2$/s \\
  \noalign{\smallskip}
  \hline
  \end{tabular}

\end{table}
\begin{figure}[th] 
\includegraphics[width=0.55\columnwidth]{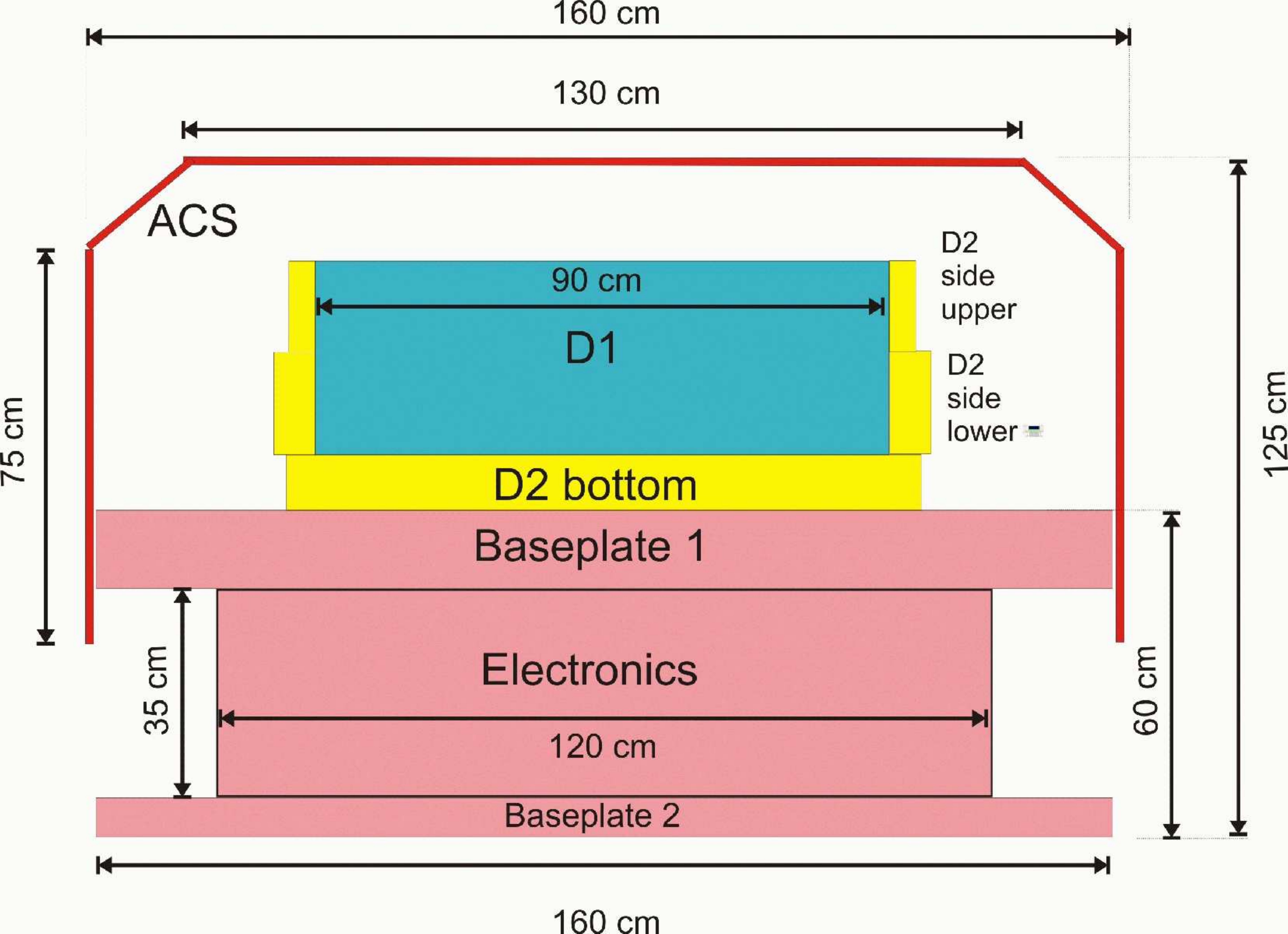} 
\includegraphics[width=0.45\columnwidth]{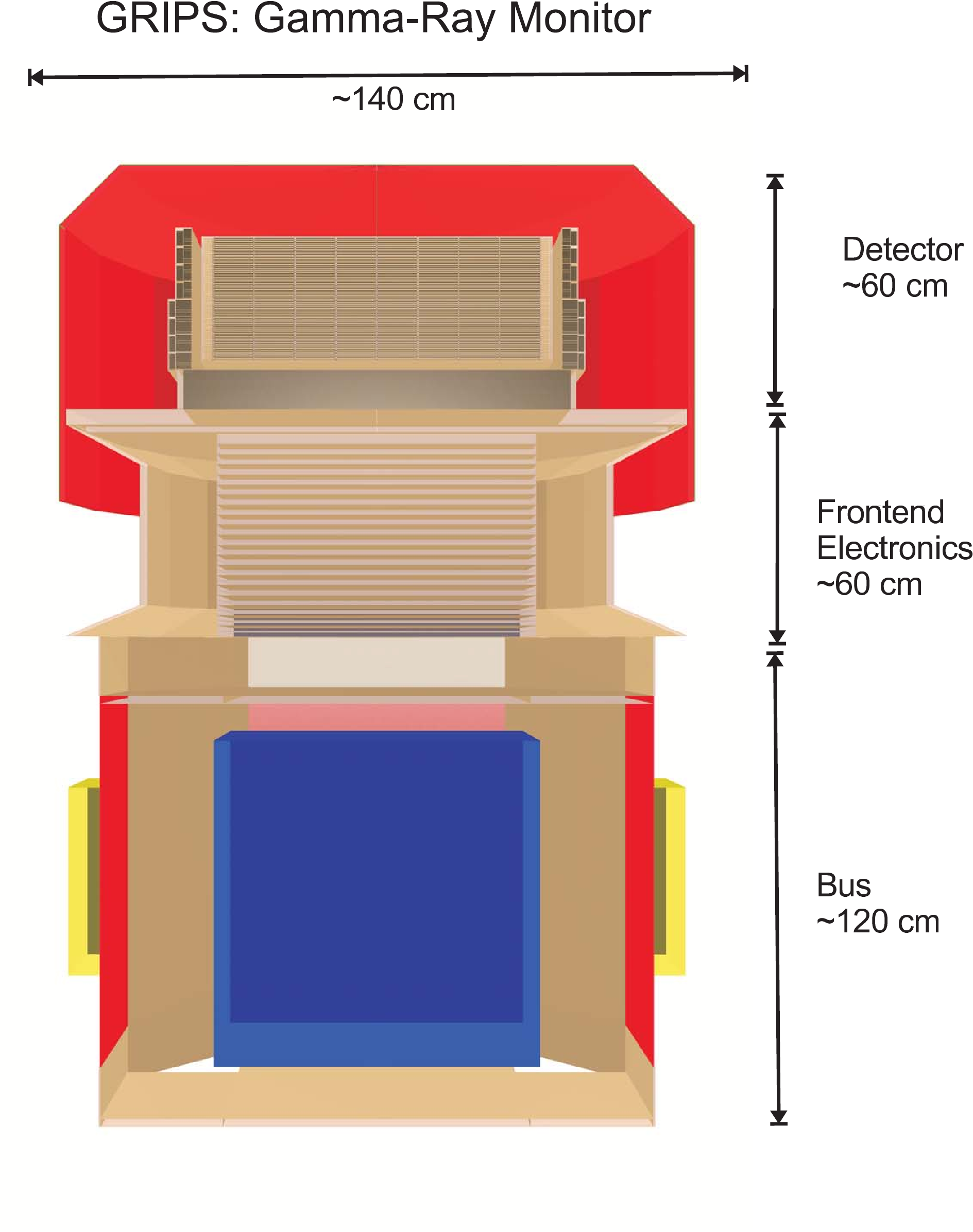} 
\caption[Size of the Gamma-Ray Monitor]{{\bf Left:} Geometry and 
size of the detector module used for the simulations.
{\bf Right:} Expected size and assembly 
of the Gamma-Ray Monitor and related electronics (top) on a  
generic satellite bus (bottom). 
\label{3D}} 
\vspace{-0.3cm}
\end{figure}

The differential Klein-Nishina cross-section for Compton 
scattering contains a strong dependence on the polarisation of the 
incident $\gamma$-ray photon. Scattered photons are emitted 
preferentially perpendicular to the direction of the electric 
field vector of the incoming photon. The strongest azimuthal 
modulation in the distribution of scattered photons will be for 
the lowest $\gamma$-ray energies and scatter angles of 
$60^\circ$-$90^\circ$. This makes a Compton telescope with a 
calorimeter covering a large solid angle a unique polarimeter.  
 
In the case of pair production, the incident photon converts into an 
electron-positron pair in the tracker. These two particles are 
tracked and determine the incident photon direction. The total 
energy is measured through the deposits absorbed in the 
tracker and/or the calorimeter.

In addition to the `telescope-mode' described  
above, the D2 detectors can also be read out in the so-called  
`spectroscopy-mode', i.e. recording interactions that deposit  
energy only in the calorimeter. Using the side walls and the  
bottom of D2 as separate units a coarse localisation and  
high-quality spectra of GRB events can be obtained. This mode of  
operation follows the examples of BATSE   
on CGRO, the ACS in INTEGRAL and the GBM on Fermi.

\subsubsection{GRM design, simulations and electronics}

The design of a new high-energy $\gamma$-ray telescope must be 
based on numerical simulations as well as experimental detector 
developments. The baseline design and input to the Monte-Carlo 
simulations of the GRM is shown in Fig. \ref{3D}. The top part 
shows the detector head with the central stack of double-sided 
Si-strip detectors (tracker D1) surrounded by the pixellated 
calorimeter (D2) and an anticoincidence system (ACS) made of 
plastic scintillator. The simulations were carried out with the 
tools of the MEGALIB software suite 
\cite{zas06,zas08}, which, 
in addition to allowing modelling of the instrument functions, 
also generates realistic background environments and traces their 
impact on the telescope for the chosen orbit. Below the 
$\gamma$-ray detector are
 the GRM electronics and a generic spacecraft bus. In 
the simulations the bus of the Advanced Compton Telescope study 
was used \cite{boggs06}. 

The D1 detector consists of 64 layers each 
containing a mosaic of $8\times 8$ double-sided Si strip detectors 
of area $10\times 10$\,cm$^2$, each of those having 96 strips per side. 
The layers are spaced at a distance of 5\,mm. 
The D2 calorimeter is made of LaBr$_3$ prisms ($5\times 
5$\,mm$^2$ cross-section) which are read out with Si Drift Diode 
(SDD) photodetectors. The upper half of the D2 side walls feature 
scintillators of 2\,cm length and the lower half has 4\,cm thick 
walls. The side wall crystals are read out by one SDD each. The 
bottom calorimeter is  8\,cm thick and is read out on both ends of 
the crystals to achieve a depth resolution of the energy deposits. 

The whole detector is surrounded by a plastic scintillator counter
that acts as an ACS against charged
particles. Read out of the scintillation light, which is collected
with embedded wavelength-shifting fibers to ensure the ACS
uniformity, will be done with Si APD detectors.

Most of the structural material that holds the detector elements
will be fabricated with carbon-fiber compounds rather than aluminum
in order to reduce the background of activation radioactivity produced 
by cosmic-rays. 
The parameters and properties 
of the detector subsystems that were used in the
simulations are listed in Table \ref{KeyChar}.

The layout of the GRM electronics is unchanged relative to our CV2007
proposal, and the interested reader is referred to \cite{grei09}.

\subsubsection{Performance}

GRIPS GRM features marked improvements when
compared to the previous generation of Compton and pair creation 
instruments. The  tracking volume (D1)
embedded in the well-type calorimeter (D2) affords a large solid angle
to detect the scattered components, which leads to a much wider FOV
than {\it COMPTEL},  and moreover to a good polarimetric response. 
Low-energy pair particles are imaged in the tracker
with much less scattering because, in contrast to the {\it EGRET} and
{\it Fermi-LAT} chambers with their metal conversion plates, the GRM has very
thin and fully active tracking detectors. 

\begin{table}[!th]
\vspace{-0.2cm}
\caption[Gamma-Ray Monitor Continuum sensitivity]{Continuum
point-source sensitivity of the Gamma-ray Monitor after 10$^6$ sec
effective exposure; $\Delta E/E \sim 1$ and comparison with previous
instruments.}
\begin{tabular}{c|c|c|c}
 \hline
 \noalign{\smallskip}
      &  \multicolumn{2}{c|}{GRIPS}             & Other$^1$      \\
      &  on-axis          & all-sky scan        & on-axis \\
  E   &  pointing         & (average)           & pointing           \\
$\!\!\!\!$(MeV)$\!\!$ & (ph/cm$^2$/s) & (ph/cm$^2$/s)   & (ph/cm$^2$/s)      \\
 \noalign{\smallskip}
 \hline
 \noalign{\smallskip}
 0.2& 4.0$\times 10^{-5}$ & 1.3$\times 10^{-4}$ & I: 1.4$\times 10^{-4}$\\
 1.0& 1.0$\times 10^{-5}$ & 2.9$\times 10^{-5}$ & C: 4.0$\times 10^{-4}$\\
 5.0& 6.5$\times 10^{-6}$ & 2.8$\times 10^{-5}$ & C: 8.0$\times 10^{-5}$\\
 20 & 1.0$\times 10^{-6}$ & 4.9$\times 10^{-6}$ & C: 1.0$\times 10^{-5}$\\
 50 & 3.2$\times 10^{-7}$ & 1.0$\times 10^{-6}$ & E: 7.8$\times 10^{-7}$\\
 80 & 2.4$\times 10^{-7}$ & 7.9$\times 10^{-7}$ & E: 2.5$\times 10^{-7}$\\
 \noalign{\smallskip}
 \hline
 \end{tabular}

  $^1$ C=COMPTEL, I=IBIS, E=EGRET.
\label{contsens}
\vspace{-0.2cm}
\end{table}

\begin{figure}[ht]
\includegraphics[width=0.55\columnwidth]{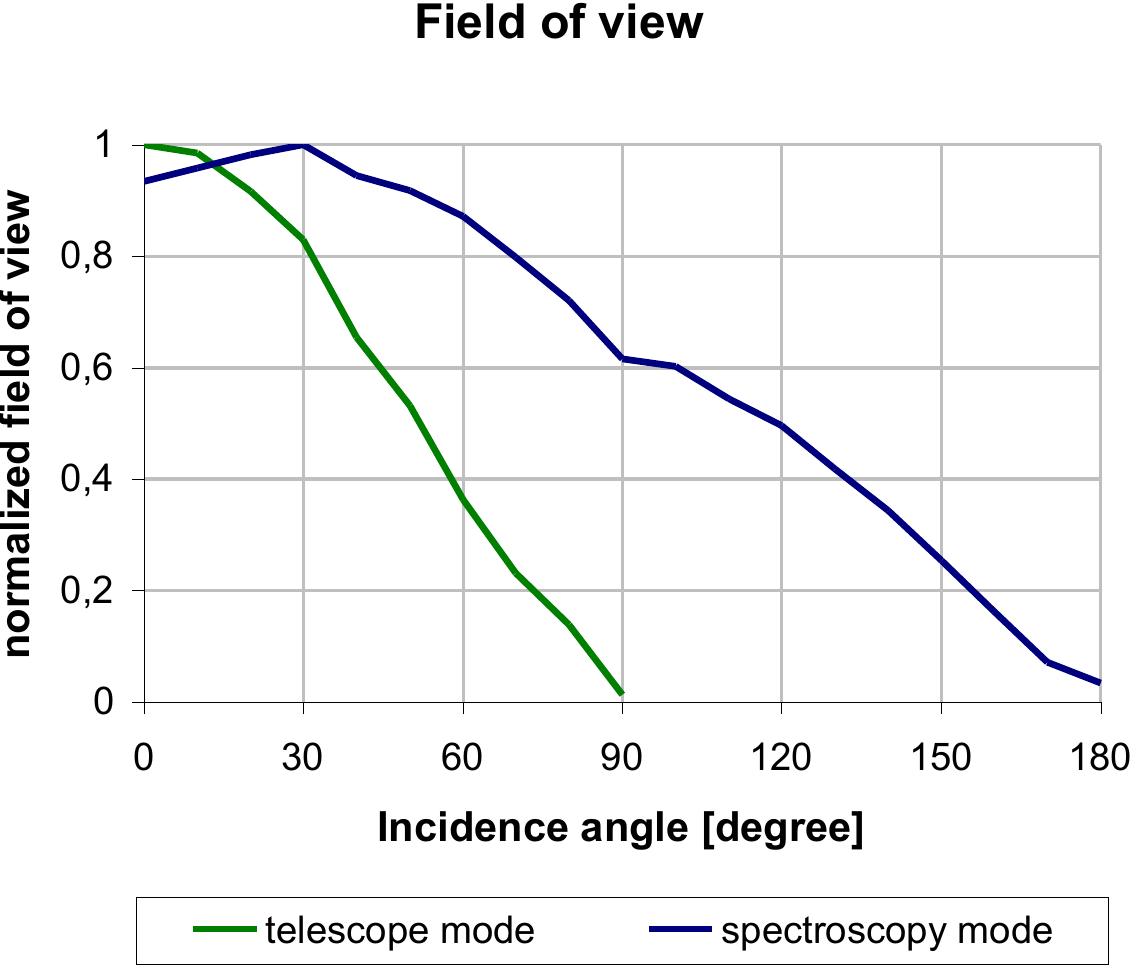}
\hfill\parbox[t]{4.cm}{\vspace*{-1.3cm}\caption[Gamma-Ray Monitor Field of view]{Field of view 
of the Gamma-Ray Monitor. In spectroscopy mode it is nearly 2$\pi$. }
\label{FOV}}
\end{figure}

These performance enhancements
can only be quantified  numerically taking into account the interaction,
detection, and reconstruction processes.
Therefore, the performance of the GRM in an equatorial low-Earth orbit
was extensively simulated with the MGGPOD suite and analyzed with the
MEGAlib package (see Zoglauer et al. 2008 for an earlier version). 
MEGAlib contains a geometry and detector description tool that was used to
set up the detailed modelling of the GRM with its detector types
and characteristics. The geometry file is then used by the
MGGPOD simulation tool 
to generate artificial events. The event reconstruction algorithms for the
various interactions are implemented in different approaches
($\chi^2$ and Bayesian). The high level data analysis tools
allow response matrix calculation, 
image reconstruction (list-mode likelihood
algorithm), detector resolution and sensitivity determination,
spectra retrieval, polarisation modulation determination etc.
Based on many billions of simulated events we have derived a
good understanding of the properties of GRIPS. 
Fine-tuning the detector concept since the CV2007 proposal has led to a 
substantial reduction in read-out channels at identical sensitivity.

{\bf Continuum sensitivity:}
GRIPS will achieve a major improvement in sensitivity over previous and 
presently active missions (Tab. \ref{contsens}), e.g. a factor 40
improvement over {\it COMPTEL} around 1--2 MeV, or a factor of $>$20 over IBIS
above 300 keV. 
The FOV (Fig. \ref{FOV})
extends to large off-axis angles: in 'telescope-mode'
up to $\sim$$50^\circ$ incidence angle for the 50\% level, or
all-sky in 'spectroscopy-mode'.

\begin{table*}[h]
\caption[]{Break-down of GRIPS improvement relative to COMPTEL at 1.8 MeV}
\vspace{-0.2cm}
\begin{tabular}{lcccc}
  \noalign{\smallskip}
  \hline
  \noalign{\smallskip}
 $\!\!$Parameter & COMPTEL & GRIPS & Improve- & Sensitivity \\
            &         &       & ment     & $\!\!$improvement$\!\!$ \\
  \noalign{\smallskip}
    \hline
  \noalign{\smallskip}
 $\!\!$Effective area (similar selection) & 16 cm$^2$  & 195 cm$^2$& 12.2 & 3.49 \\
 $\!\!$Observing time (pointing vs. scanning) & 0.35 & 1   & 2.9  & 1.69 \\
 $\!\!$Energy resolution                   & 59 keV  & 17 keV  & 3.5  & 1.87 \\
 $\!\!$Angular resolution                  & 3\fdg9  & 1\fdg5  & 2.6  & 1.61 \\
 $\!\!$Field-of-View (HWHM)           & 30\degs\   & 45\degs  & 2.2  & 1.48 \\
 $\!\!$Background (orbit, passive material, $\!\!$  & & & 3.0 & 1.73 \\
        tracker shielding)                       & & & & \\
    \hline
 $\!\!$ Total                         &      &     &      & 45.2 \\
    \hline
  \end{tabular}
  \label{factors}
  \vspace{-0.1cm}
\end{table*}

\begin{table}[ht]
\vspace{-0.2cm}
\caption[Gamma-Ray Monitor Narrow line sensitivity]{Narrow-line
point-source sensitivity of the GRM after 10$^6$ sec
effective exposure.}
\vspace{-0.1cm}
\begin{tabular}{c|c|c|c}
 \noalign{\smallskip}
 \hline
 \noalign{\smallskip}
      &  \multicolumn{2}{c|}{GRIPS}             & SPI                \\
      &  on-axis          & all-sky scan        & on-axis            \\
  E   &  pointing         & (average)           & pointing           \\
$\!\!$(keV) & ~(ph/cm$^2$/s)~ & ~(ph/cm$^2$/s)~   & ~(ph/cm$^2$/s)     \\
 \noalign{\smallskip}
 \hline
 \noalign{\smallskip}
~511& 3.4$\times 10^{-6}$ & 1.5$\times 10^{-5}$ & 5.1$\times 10^{-5}$\\
1157& 2.4$\times 10^{-6}$ & 7.8$\times 10^{-6}$ & 3.0$\times 10^{-5}$\\
1809& 1.5$\times 10^{-6}$ & 4.1$\times 10^{-6}$ & 3.2$\times 10^{-5}$\\
 \noalign{\smallskip}
 \hline
 \end{tabular}
\label{narlinsens}
\vspace{-0.25cm}
\end{table}

{\bf Line sensitivity:}
The narrow-line point-source sensitivity
of the GRM for three astrophysically important $\gamma$-ray lines
is given in Tab.\,\ref{narlinsens}. 
For the standard operation
mode, the all-sky scan, the resulting sensitivity is slightly worse than
in pointing mode since the exposure is distributed over most of the sky
and not concentrated into a small FOV like for INTEGRAL. This
reduction is, however, offset by the large geometrical factor (effective
area $\times$ solid angle), the uniformity of the scan, and the
permanent accumulation of exposure. As a consequence, after 5 years in
orbit, GRIPS will achieve a factor 40 sensitivity improvement over
{\it COMPTEL} (in 9 yrs) in e.g. the 1809 keV line.
A breakdown into the various
factors is shown in Tab. \ref{factors}, where we have assumed (there is
no way to compare this directly) that the different measurement techniques, 
i.e. tracking and multiple Compton interaction in GRIPS vs. time-of-flight
and PSD selection in {\it COMPTEL}, have about the same efficiency.

\begin{figure}[hb]
\vspace{-0.4cm}
\includegraphics[width=0.52\textwidth]{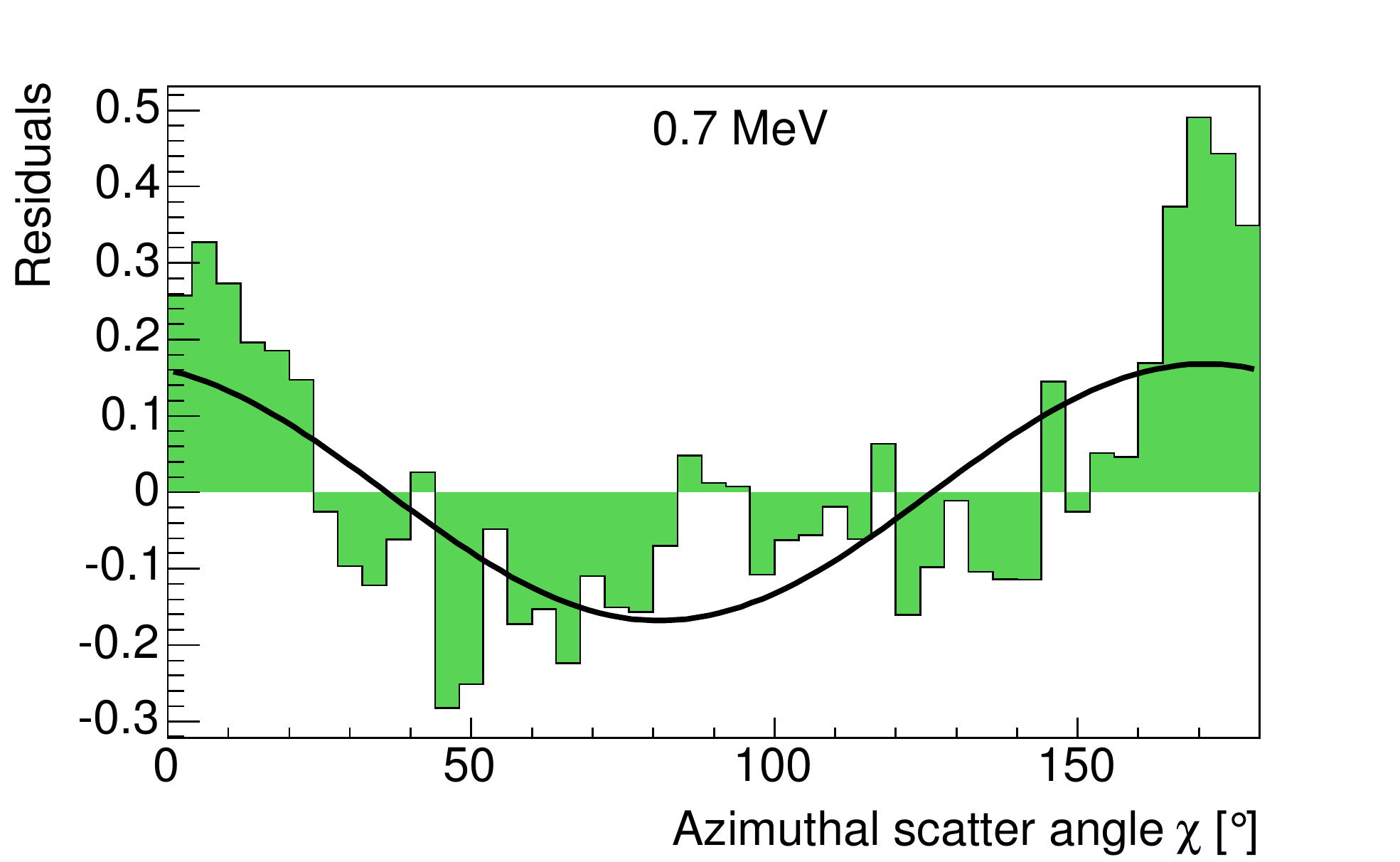}
\includegraphics[width=0.52\textwidth]{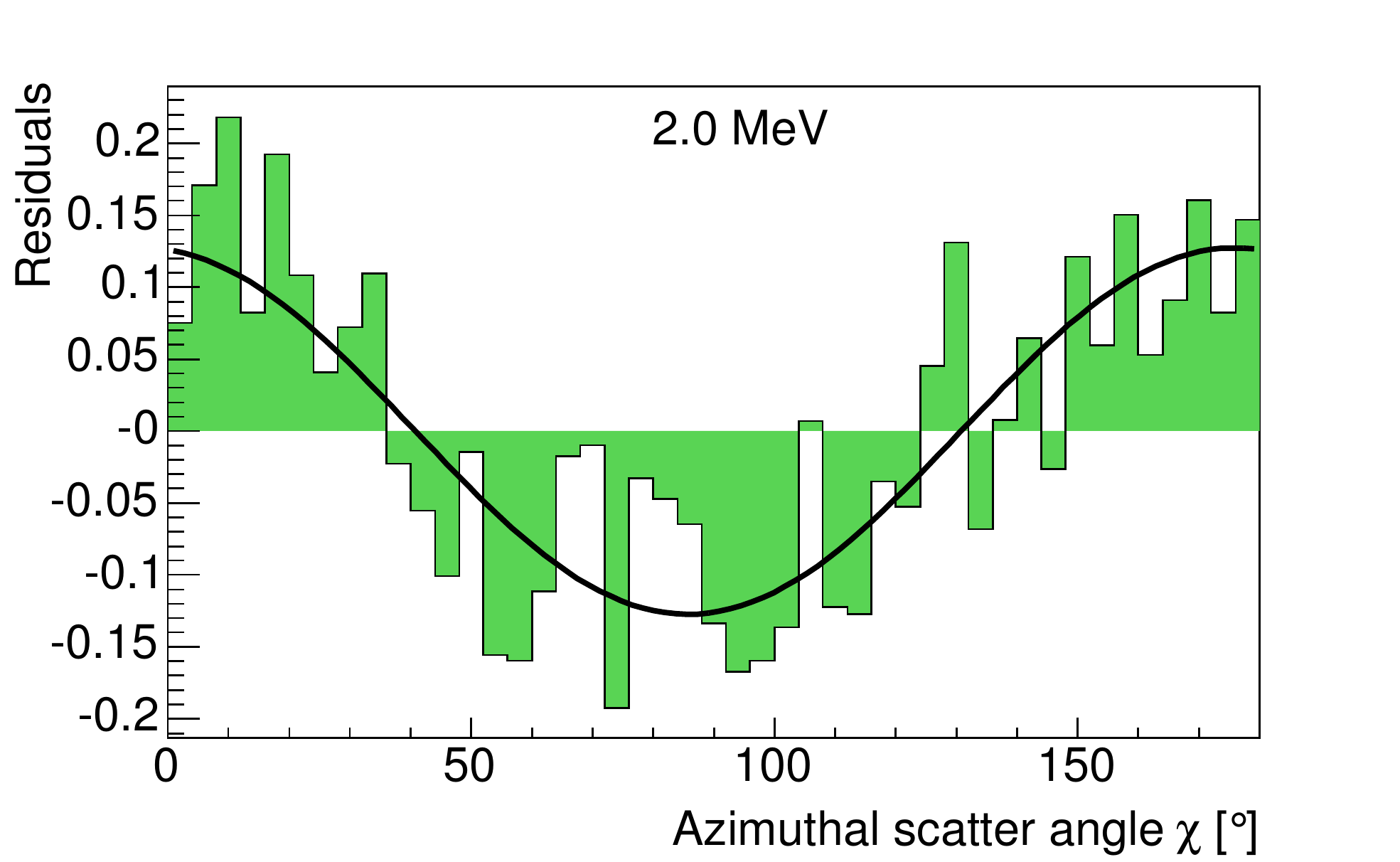}
\caption[Polarisation response of MEGA]{Measured polarisation response of
the MEGA prototype for two different energies. Within measurement
errors and statistics, all values are in agreement with GEANT4 
simulations: the polarisation angle of 90\degs\ is reproduced
to 82\%$\pm$24\% (0.7 MeV) and  86\%$\pm$11\% (2.0 MeV), respectively.
The measured modulation is 0.17$\pm$0.04 and  0.13$\pm$0.03
 as compared to the simulated values 0.19 (at 0.7 MeV) and 0.14 (2.0 MeV).
(From \cite{zog05}) \label{mepola}}
\vspace{-0.2cm}
\end{figure}

{\bf Polarisation:}
Linearly polarised $\gamma$--rays
preferentially Compton scatter perpendicular to the incident polarisation
vector, resulting in an azimuthal scatter angle distribution
(ASAD) which is modulated relative to the distribution for
unpolarised photons. The sensitivity of an instrument to
polarisation is
given by the ratio of the amplitude of the ASAD and its average,
which is called the modulation factor $\mu$. The modulation
is a function of incident photon energy, \emph{E,} and the Compton
scatter angle, $\theta$, between the incident and scattered photon
directions (see Fig. \ref{mepola} for the MEGA prototype calibration).

\begin{figure}[th] 
\includegraphics[width=0.6\textwidth]{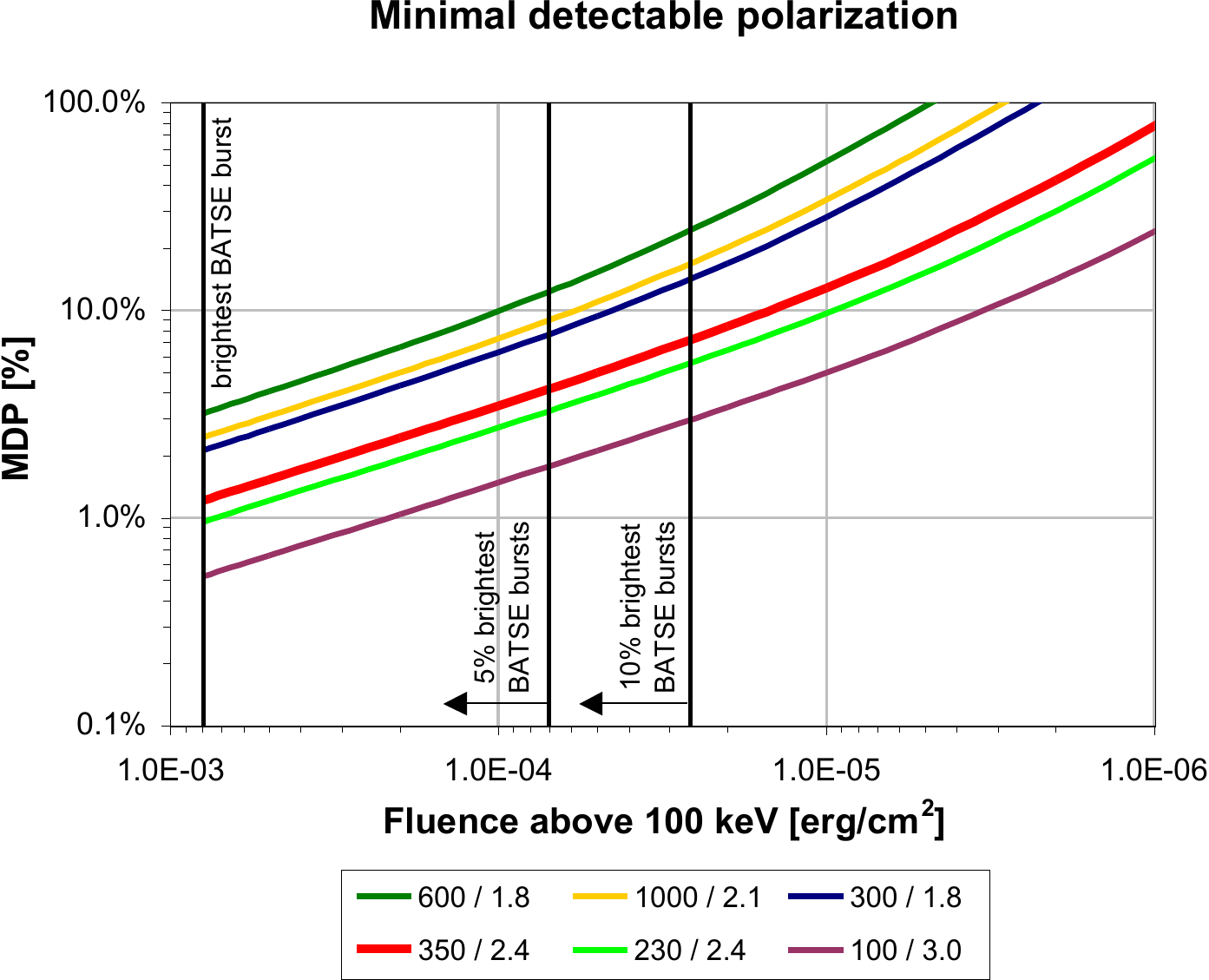} 
\hfill\parbox[t]{4.cm}{\vspace*{-3.7cm}\caption[Gamma-Ray Monitor Polarisation sensitivity]{Polarisation 
sensitivity  of GRM.  
Various models of GRB spectra are shown 
which differ in their break energy  and their high-energy power law slope; 
see legend for the parameter pair for each model. 
Note that at the bright end the minimal detectable polarisation  
changes much more slowly than the fluence. }
\label{polasens}} 
\vspace{-0.2cm} 
\end{figure}

GRIPS is a nearly perfect polarimeter (Fig. \ref{polasens}): 
The well-type geometry
allows the detection of Compton events with large scatter angles which
carry most of the polarisation information. 
The best polarisation sensitivity is achieved in the 200--400 keV range.

{\bf Gamma-ray Bursts:} We have created model spectra with
parameters for $E_{\rm peak}$, high-energy power law slope $\beta$
and peak flux, which cover their distribution  as
observed with BATSE (4th BATSE catalog; Paciesas et al. 1999). Note that
the GBM/Fermi distribution of these parameters
is, within statistics, identical to those of BATSE despite the wider energy 
range of GBM, thus indicating that the parameter space is complete. 
The faintest BATSE GRBs are clearly `detected' by GRIPS with spectra
extending from 100\,keV up to 2\,MeV and having more than 5 energy
bins with 3$\sigma$ each. From these simulations we estimate that
GRIPS will be a factor of 3 more sensitive than BATSE below 1\,MeV
and detect about a factor of 2.5 more GRBs than BATSE. Folding in
the smaller field of view of GRIPS compared to BATSE, we find that
GRIPS will detect 665 GRBs/yr. For GRBs at an off-axis angle of
$>$70\degs, only poor localisations, if any, will be possible, so
we expect 440 GRBs/yr with good positions and XRM follow-up.

The present Swift samples of GRBs, both large biased samples as well as 
smaller but nearly complete samples \cite{gkk10}, 
indicate a fraction of 5.5$\pm$2.8\% GRBs at z$>$5. 
Using standard cosmology and star formation history description, this 
translates into a fraction of 1\% of all GRBs located
at z$>$10, or 0.1\% of all GRBs at z$>$20. With 440 GRBs per
year, and a nominal lifetime of 5 yrs (goal 10 yrs) we would
expect 22 (goal 44) GRBs at z$>$10, and 2 GRBs (goal
4) at z$>$20. 
This includes a duty
cycle of the instrument similar to that of BATSE and a detection
rate of X-ray afterglows of 98\%, as for Swift/XRT. 
Measurements with the IRT will ensure that high-z (z$>$7) candidates are
flagged immediately, and thus receive special attention in the 
optical/NIR identification and spectroscopic follow-up.
The above numbers imply that
(1) the detection frequency of GRBs at z$>$20 is high enough to
achieve at least one detection during the mission lifetime, and
(2) that GRIPS will clearly detect the cut-off in the
$z$-distribution IF star formation starts at a certain redshift
(below $\sim$25) throughout the Universe. 
Measuring this cut-off, or no cut-off up to
$z$$\sim$25, would in turn be a limit to the earliest time when
stars formed.

\begin{figure}[th]
\vspace{-0.2cm}
\includegraphics[width=\textwidth,]{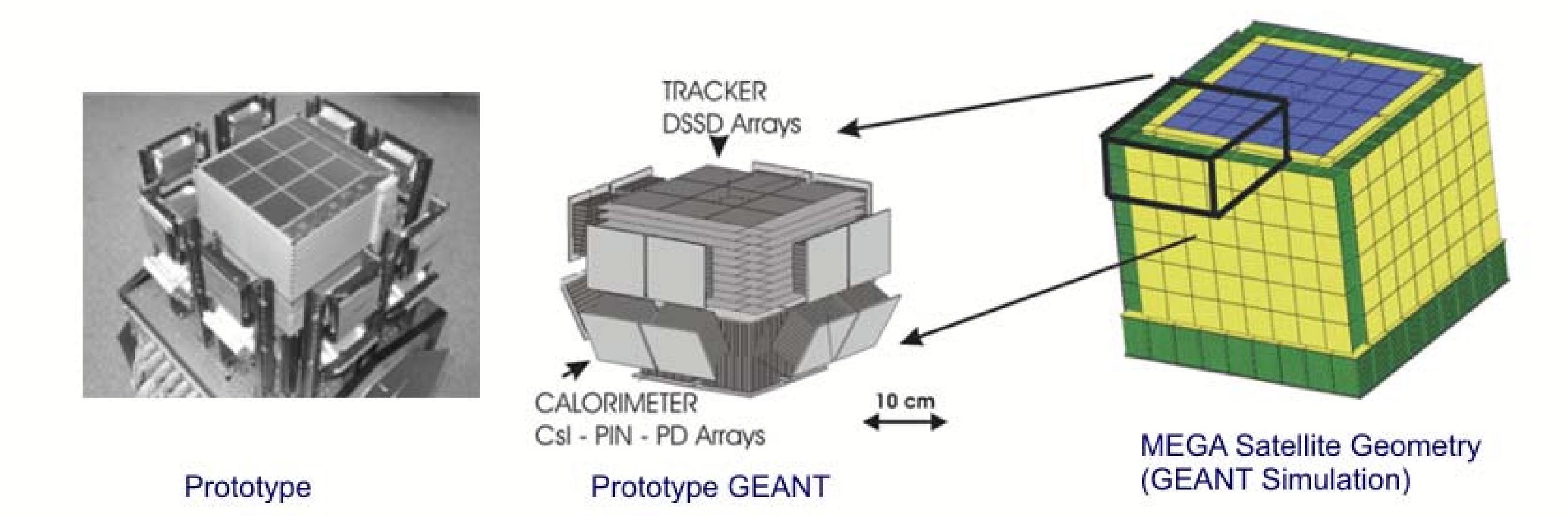}
\vspace{-0.3cm}
\caption[GEANT model of MEGA]{The GEANT models  for the MEGA satellite version
(right), prototype (middle) and a photograph of the actual
prototype detector (open to show the DSSD and CsI detectors).}
\label{MEGA_proto}
\end{figure}

{\bf Number of expected sources:} 
{\em Extrapolation from Swift/BAT (15--150\,keV):}  
Since this requires the assumption that the spectra do not cut off at 
some intermediate energy,  
we base our extrapolation on the 
10\% fraction of blazars (out of the total AGN population) 
for which there is no spectral break. 
We assume conservatively a photon index 
of -1.5 and using the logN-logS identified during the first three 
years of Swift/BAT \cite{agk07}, 
we find that GRIPS 
will detect, in a 1 year  exposure, about 820 blazars in the 
1--10\,MeV band, among those 4 at $z>8$.

{\em Extrapolation from Fermi/LAT:}
Fermi/LAT located 1451 sources during its first year survey (1FGL).
About 700 objects are associated with extragalactic sources and include 
blazars, Seyfert and starburst galaxies. Among the identified
galactic sources about 60 young pulsars and PWNe and 20 millisecond pulsars
stand out, but many low galactic latitude sources are also associated with 
SNRs and a few HMXBs were detected. 
Detection or non-detection of all 1FGL sources
in the adjacent
lower energy range will be extremely important to understand their radiation
processes.
We estimate the detectability of the 1FGL sources in the 1--50 MeV range by
extrapolating the power-law spectra measured by the LAT above 100 MeV,
with the caveat that the spectrum could turn over
(many EGRET-COMPTEL correlations showed spectral breaks in the 1-10 MeV range).
We use not only the
1FGL spectral index (SI) but also the uncertainties, ie softer 
(S.I. + error S.I.) or harder (S.I. -
error S.I.) spectral indices.  
For the 1--10 MeV range we 
find that between 860 and 1200 1FGL sources are detectable (Tab. \ref{srcnum}), 
and  between 420 and 740 in the 10--50 MeV range.

{\em Predictions from theory/modelling:}
The number of new source
types in the ``discovery space'' domain is more difficult to estimate.
Based on theoretical predictions of emission properties (dominant energy band,
fluxes) and folded with the GRIPS sensitivity and the all-sky survey mode
yields the following numbers: 
wind-collisions in high-mass binaries (200 keV): 30;
superbubbles/star-forming regions (0.5-7 MeV): 20-200;
microquasars/BH transients (200 keV/511 keV): 2/yr;
supersoft sources (2.2 MeV): 10;
flare stars (200 keV): 10/yr;
galaxy cluster (4-7 MeV): 50.

\begin{table}[th] 
\vspace{-0.2cm} 
\caption[Number of expected sources]{Number of expected sources 
of various classes after 1 yr and 5 yrs of GRIPS exposure. 
The last column gives the number of completely new sources, not known 
before in any other wavelength. In addition, 
about 1500 steady sources known in other wavelength bands  
will be detected in the MeV band for the first time. \label{srcnum}} 
\vspace{-0.2cm} 
\begin{tabular}{lrrc} 
  \noalign{\smallskip} 
  \hline 
  \noalign{\smallskip} 
  Type           & 1 yr & 5 yrs & new \\ 
  \noalign{\smallskip} 
  \hline 
  \noalign{\smallskip} 
  GRBs                     & 660 & 3300 & 3300 \\ 
  Blazars                  & 820 & 2000 & 400 \\ 
  Other AGN (100-300 keV)~~ & 250 & 300 & 0? \\ 
  Pulsars/AXP              &  30  & 50 & 0? \\ 
  Supernovae (Ia and cc)   &   2  & 20 & 20 \\
  Unidentified sources     & 170 & 230 & 60 \\ 
  \hline 
  \end{tabular} 
\vspace{-0.5cm} 
\end{table}

\subsubsection{Operation}

The GRM instrument is highly pixelated and features about  $3\times 10^5$
 fast measurement channels. Thus, the power ($\approx$300 W), 
cooling ($\approx$100 W) and data handling
resources are large, but not challenging.
According to our design, the electronics will be placed in a
separate compartment from the detector units (see Fig. \ref{3D}).
Only a pre-amplifier and line driver will be integrated directly
on the detectors, allowing the signals to be conducted to the
remote (1--1.5\,m distance) electronics. In this way the heat
dissipation in the very densely packed detector unit will be
minimised and the operating conditions of around 10\degs C can be
established by passive thermal maintenance. 
The overall thermal layout, one of the major challenges of our CV2007
proposal \cite{grei09}, has been drastically mitigated by reducing the 
number of channels.

GRM will generally be operated in a continuous zenith pointed
scanning mode. The field of view (diameter 160$\degs$) will cover
most of the sky over the course of one orbit, similar 
to the all-sky survey performed by LAT
on {\it Fermi}. Since the XRM/IRT are on a separate 
satellite ($\S$ 6), their frequent pointing changes have no impact on the
GRM survey.

\subsection{X-ray Monitor}

The main driver for the design of the X-ray monitor (XRM) is the 
positional accuracy of GRBs, so that their full error circle 
can be covered by the XRM. GRM will determine GRB positions
to better than 1\degs\ (radius, 3$\sigma$) down to off-axis
angles of 60\degs. We add a 50\% margin, and require a field
of view of the XRM of 3\degs\ diameter.

The second requirement is for sensitivity which should be at least
a factor 3 larger than that of {\it Swift}'s XRT, since GRIPS will
cover more distant GRBs, and thus likely fainter afterglows.
Such sensitivity requirement (of order $>$300 cm$^2$) excludes
coded mask systems, and even single-telescope Wolter-I 
optics are problematic.

\subsubsection{Measurement technique, Design and Key characteristics }

\begin{figure}[th]
\hspace{-0.2cm}
\includegraphics[width=0.54\columnwidth]{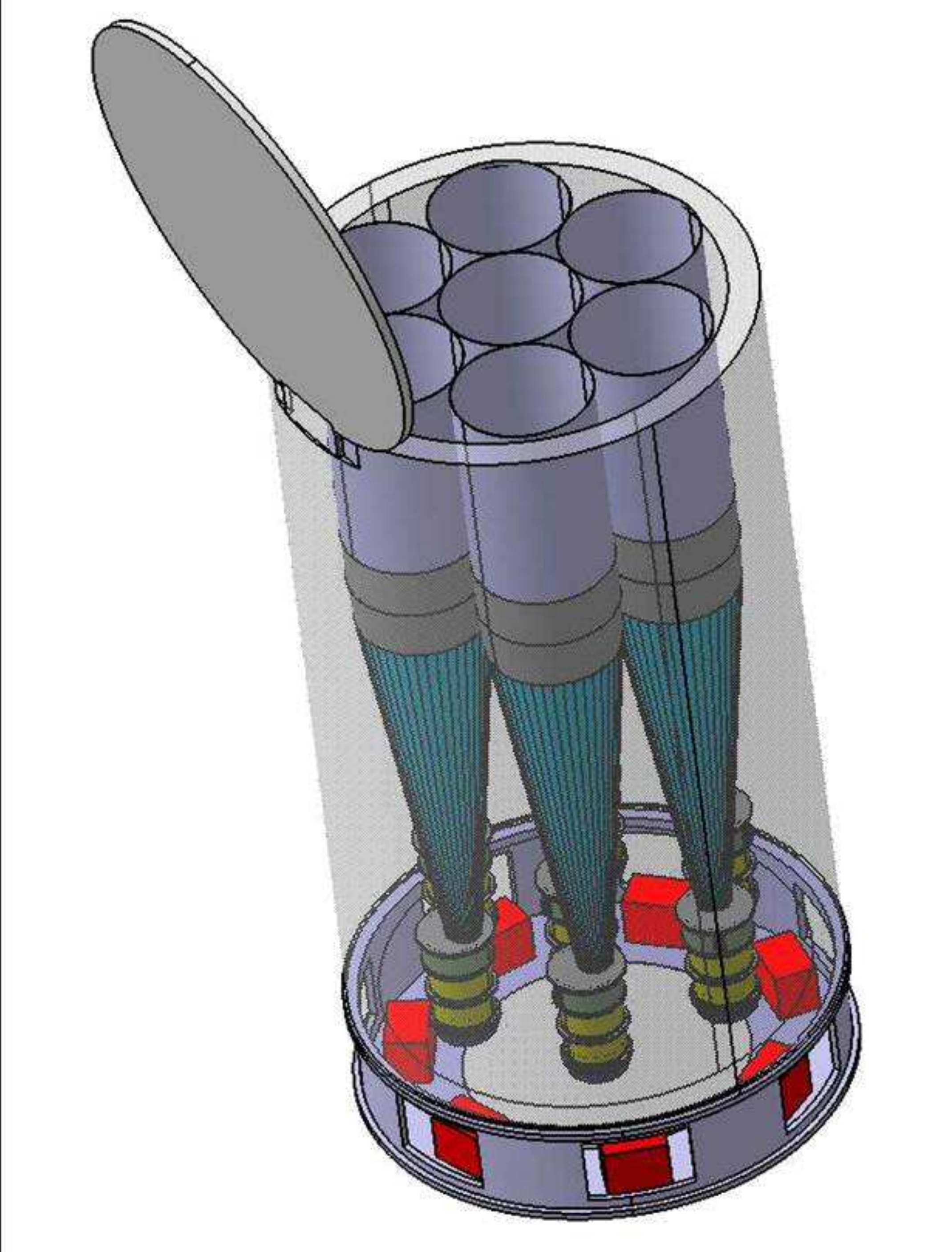}
\includegraphics[width=0.4\columnwidth]{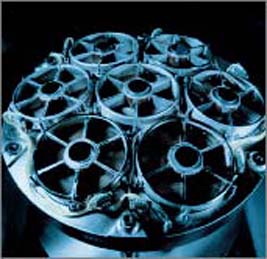}
\vspace{-0.2cm}
\caption[eROSITA Telescope]{Scetch of the 7-telescope configuration
(left) and the prototype telescope array mounting (right) 
of eROSITA  which will be adapted for the XRM of GRIPS.}
\label{XRMTel}
\end{figure}

We therefore embark on a design consisting of multiple Wolter-I telescopes.
The easiest and probably most cost-effective option is to
adopt the {\it eROSITA} scheme of 7  Wolter-I telescopes (Fig. \ref{XRMTel}), 
and adjust their
orientation on the sky such that they fill 
the required FOV.
{\it eROSITA} (for {\bf e}xtended {\bf RO}entgen {\bf S}urvey with an 
{\bf I}maging {\bf T}elescope {\bf A}rray) shall perform the first 
imaging  sky survey in the medium X-ray range, i.e. between 0.2 and 12 keV,
with a sensitivity of 9$\times$10$^{-15}$ erg cm$^{-2}$ s$^{-1}$ (0.2-2 keV)
and 2$\times$10$^{-13}$ erg cm$^{-2}$ s$^{-1}$ (2--10 keV) \cite{predehl2010}. 
This will allow the detection of about 3.2 million AGN, and  
$\approx$100.000 cluster of galaxies.
The satellite will
be launched with a Soyus-Fregat rocket from Baikonur.
The {\it eROSITA} instrument development is led by MPE Garching, and the main
key characteristics are summarized in Tab. \ref{xrmkey}.

\subsubsection{Performance}

\begin{figure}[hb]
\includegraphics[width=0.65\columnwidth]{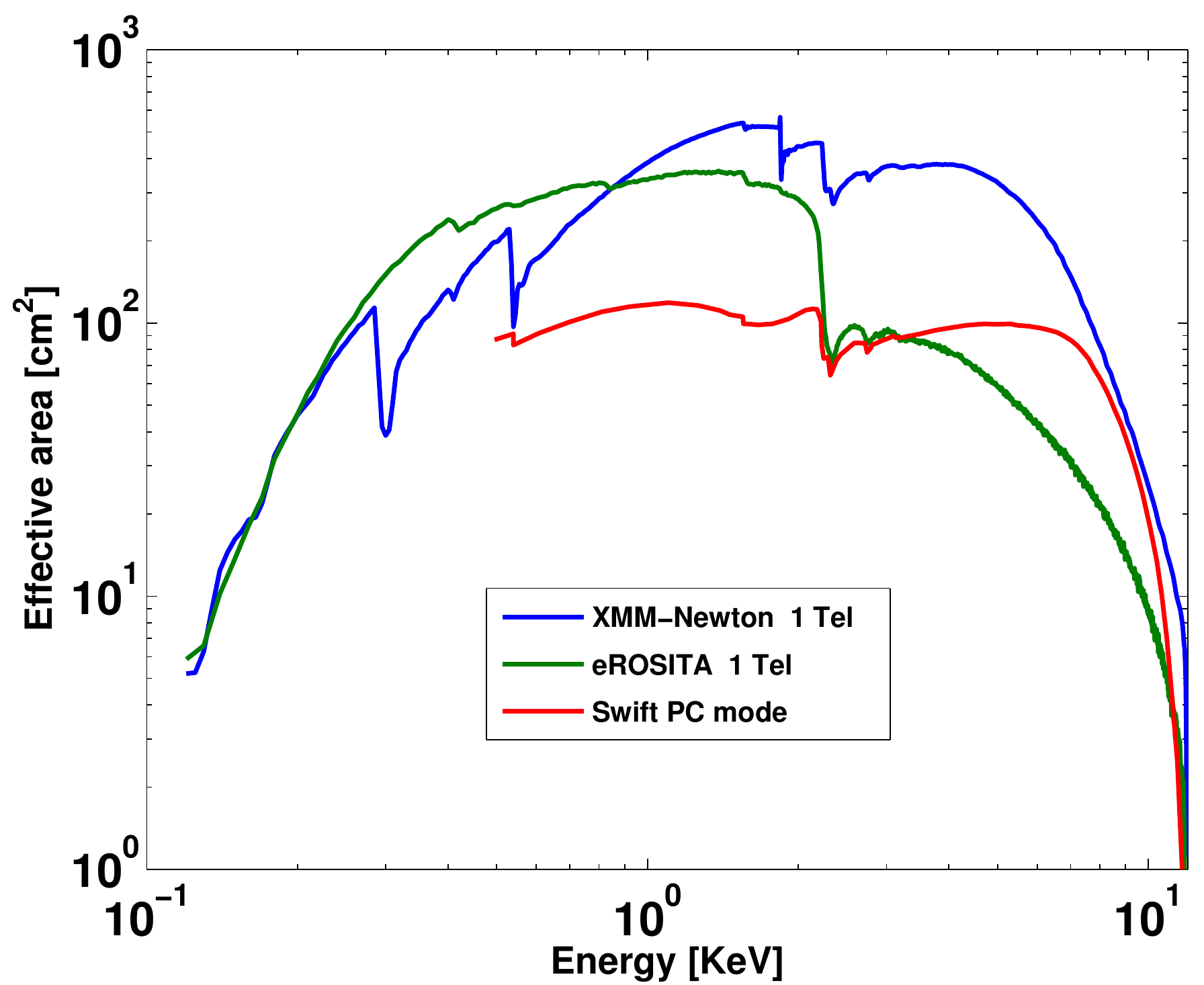}
\vspace{-0.3cm}
\hfill\parbox[t]{3.5cm}{\vspace*{-2.5cm}\caption[XRT effective area comparison]{Comparison of the effective
area of the modified eROSITA system (one telescope per sky position)
with those of XMM and Swift/XRT.
\label{xrteffarea}}}
\vspace{-0.3cm}
\end{figure}

\vspace{-0.2cm}
We mention a number of improvements of the {\it eROSITA}
detector over the {\it XMM}-Newton EPIC/PN detector: 
(i) lower noise and very homogeneous over detector pixels,
(ii) smaller charge transfer losses,
(iii) higher energy resolution, especially for low X-ray energies ($<$1keV),
(iv) smaller number of out-of-time events due to frame-store technique,
(v) higher time resolution of 50 ms.
As with the {\it XMM}-Newton EPIC/PN, the {\it eROSITA} detectors will share 
the high quantum 
efficiency ($>$90\% over 0.2--12 keV band), the high radiation hardness 
against high energy protons including self-shielding against the low 
energy protons focused by the Wolter telescope.
The effective area of the {\it eROSITA} telescope, even after placing all 
7 telescopes at different sky positions, well matches the GRIPS requirements
(Fig. \ref{xrteffarea}).

\begin{table}[hb]
\vspace{-0.2cm}
\caption[XRM key characteristics]{Key characteristics of the XRM and 
detector (eROSITA Design Doc; MPE 2006).}
\begin{tabular}{ll}
  \hline
  \noalign{\smallskip}
Number of mirror modules   &     7  \\
Degree of nesting          &     54 \\
Focal length               &    1600 mm \\
Largest mirror diameter    &     360 mm \\
Smallest mirror diameter   &      76 mm \\
Energy range               &     $\sim$0.2--12 keV \\
Field-of-view              &             61\amin\ $\oslash$ \\
Angular resolution (HEW)   &  28\asec\ (FOV-averaged) \\
effective area (single telescope)        &  330 cm$^2$ at 1.5 keV / 20 cm$^2$ at 8 keV \\
Mass per module            &     68 kg \\
  \noalign{\smallskip}
  \hline
  \end{tabular}
\label{xrmkey}
\end{table}

\begin{table}[ht]
\caption[XRM Resources]{XRM resources}
\begin{tabular}{ll}
  \hline
  \noalign{\smallskip}
total eROSITA mass (incl. margin)    & 635 kg \\
Power eROSITA electronics            & 125 W \\
Power mirror /telescope heating      & 140 W \\
Telemetry Events (average, 7 tel.)    & 7.0 kbit/s \\
Telemetry HK (average, 7 tel.)    & 3.6 kbit/s \\
On-board data storage (per day)     & 128 Mbyte \\
  \noalign{\smallskip}
  \hline
  \end{tabular}
\end{table}

If a $\gamma$-ray burst is detected, the GRM
satellite will autonomously issue a slew of the XRM/IRT satellite 
onto the target. 
To control the XRM/IRT pointing, two star-sensors will 
be mounted on each of the telescope structures. The autonomous pointing 
strategy has been shown on the {\it Swift} mission to be of very high 
scientific interest.

{\it eROSITA} is a fully funded project in collaboration with Russia,
scheduled for launch in 2013. 
For a planned launch of the M mission
in 2020--2022, instrument implementation will start around 2014/15,
i.e. after
completion and launch of {\it eROSITA}.
Thus, not only will all technological problems be solved,
but also all lessons learned during the development and assembly
of {\it eROSITA} can be incorporated into the design of the XRM.

\subsection{Infrared telescope}

The main driver for the design of the Infrared Telescope (IRT) is the 
goal to rapidly identify those GRBs which are at high redshift, 
so that particular care 
can be devoted to their non-GRIPS follow-up observations. After the
slew to a GRB, the XRT will provide a position with an accuracy
between 15-50\asec\, depending on the off-axis angle of the GRB in the XRT 
FOV. This uncertainty is too large for immediate (low-resolution)
spectroscopy, so we resort to imaging.

\subsubsection{Measurement technique, Design and Key characteristics }

We propose simultaneous multi-band photometry in seven channels,
zYJHKLM  to determine photometric redshifts of GRBs (Fig. 3). 
Since GRB afterglow spectra are simple power laws, and at $z>3$ Lyman-$\alpha$ 
is {\sc the} 
dominant spectral feature, relatively high accuracies can be reached 
even with broad-band filters (Fig. \ref{IRTaccuracy}), 
as demonstrated in ground-based observations with GROND \cite{ksg10}.

Since the $LM$ band afterglow detections require a 1\,m class telescope
(see below), the most cost-effective way is a copy of the telescope as
presently proposed for the EUCLID mission, but with much reduced
requirements for the tolerances in the alignment of the mirrors,
in particular the PSF ellipticity, and
long-term stability. We have replaced the EUCLID
instrumentation with a system of dichroics which split the beam into
the seven passbands, very similar to the GROND concept \cite{gbc08}.
Basic characteristics are given in Tab. \ref{irtpar}.

The baseline detector is a 2048$\times$2048, 18 $\mu$m pixel, 2HRG detector
from Teledyne Imaging Sensors. Depending on the channels, different cut-off 
wavelengths of 1, 2.5 and 5 $\mu$m will be chosen, respectively.
The read-out, analog-to-digital conversion as well as first image
processing is done by SIDECAR (System for Image Digitalization, Enhancement, 
Control and Retrieval) ASICs, also available from Teledyne.
Control electronics and post-processing/analysis CPU will be similar, but 
much simpler due to a substantial reduction in the number of moving parts 
(thus less
control electronics, drivers, position sensors) than that presently designed 
for EUCLID.

\subsubsection{Performance}

Based on a complete sample of GRB afterglow measurements obtained with
GROND since 2007 \cite{gkk10}, in particular the afterglow brightness
distribution in each of the g'r'i'z'JHK channels, we derive a minimum
afterglow brightness of $M$(AB) $\approx 22$ mag at 2 hrs after the GRB.
Using standard parameters for the transmission of the optical components,
read-out noise of the detector as well as zodiacal background light,
a 1\,m class telescope is needed to reach a 5$\sigma$ detection with
a 500 sec exposure.

\begin{figure}[hb]
\includegraphics[width=0.6\textwidth]{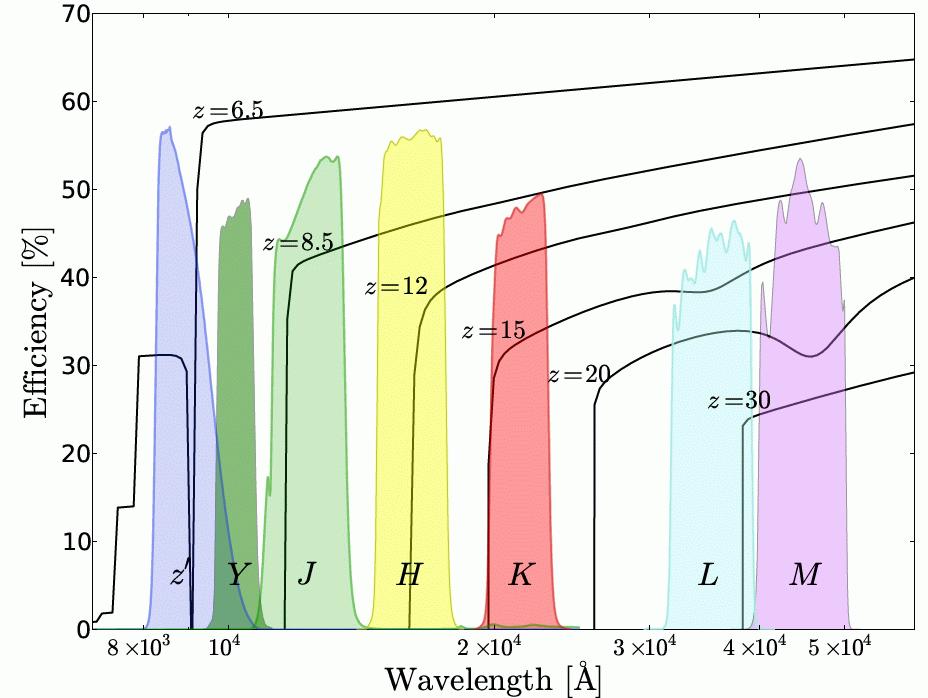}
\hfill\parbox[t]{4.5cm}{\vspace*{-5.7cm}\caption{Scetch of the proposed filter 
bands zYJHKLM for the IRT. 
The efficiencies include best-effort estimates of all
optical components, including the telescope, dichroics, filters and 
detector. Shown in black lines are template afterglow spectra, from 
redshifts 
$z$=6.5, 8.5, 12, 15, 20 to 30 (Top left to bottom right). These spectra also 
differ in their spectral index (z=6.5 vs. all others) and 
rest-frame extinction (amount and reddening law; e.g. z=8.5
and z=20, the latter of which shows a redshifted 2175\AA~dust feature. }}
\label{filters}
\end{figure}

\begin{figure}[ht]
\includegraphics[angle=90, width=0.68\textwidth]{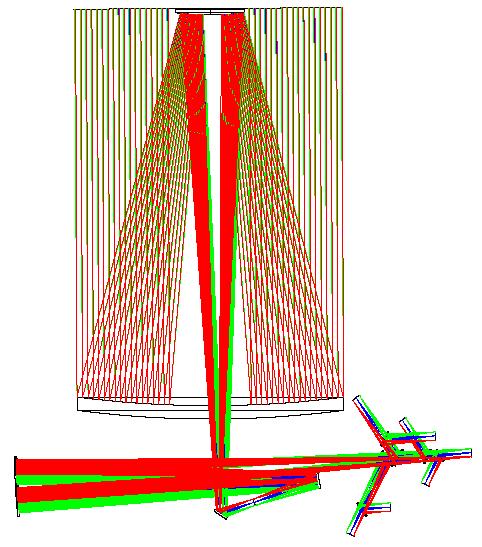}
\includegraphics[width=0.30\columnwidth]{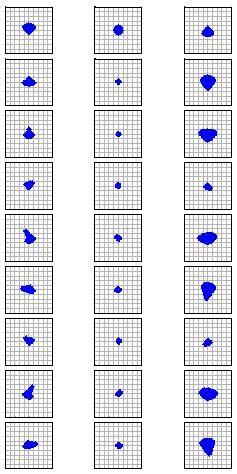}
\caption{{\bf Left:} Basic concept of the IRT with its 7 photometric channels.
M1 (diameter, curvature, but not conic constant) as well as the distance 
between M1 and M2 are identical to the EUCLID concept. M3 to M5 just serve
as flat folding mirrors to minimize the size.
{\bf Right:} Spot diagram of our ZEMAX design, for channels 
YJ  (pass through one dichroic, left),
z (only reflections, center), and
M (pass through four dichroics, right).
The top row is for the center of the FOV, and the other 8 rows are for
the sides and corners of the 10\amin$\times$10\amin\ FOV.
 The large boxes are 18 $\mu$m on a side, corresponding to one H2RG pixel, and
 the small grid is 1.8 $\mu$m.}
\label{optics}
\vspace{-0.2cm}
\end{figure}

A ZEMAX design with the EUCLID baseline (a 3-mirror concept similar
to DIVA/FAME/Gaia) and including all 7 channels 
and 3 folding mirrors returns a perfect imaging quality with a Strehl ratio
of 99\% 
(right panel of Fig. \ref{optics}) 
even for the worst channel (light passes through four dichroics).

Similar to the GROND case \cite{ksg10}, extensive Monte-Carlo simulations  
have been performed and demonstrate the accuracy with which the redshift 
can be determined (Fig. \ref{IRTaccuracy}). 
The sample properties are chosen to be as close as possible to what is  
known about optical/NIR afterglows with respect to their spectral indices, 
neutral hydrogen column densities of their DLAs, and dust extinction. 
For the GROND case and $3<z<8.2$ range, these simulations are nicely 
confirmed by bursts which have also spectroscopic redshifts. 

Estimates on the telescope have been adopted from the EUCLID Assessment 
Study Report (ESA/SRE-2009-2). Estimates on optics and detectors are
based on GROND experience.

\begin{table}[hb]
\vspace{-0.3cm}
\caption[IRT Key characteristics]{IRT key characteristics \label{irtpar}}
\begin{tabular}{cc}
    \hline
      \noalign{\smallskip}
    Parameter & Value \\
     \noalign{\smallskip}
    \hline
      \noalign{\smallskip}
    Telescope   & 1.2\,m Korsch \\
    Filter      & zYJHKLM (simultaneous)\\
    FOV         & 10\amin\ $\times$ 10\amin\  \\
    Detectors   & 7 2K$\times$2K HAWAII \\
    Plate Scale & 0\farcs3/pixel \\
    Sensitivity (5$\sigma$, 500 sec exp) & 24/23 mag AB (z-K/LM) \\
    \noalign{\smallskip}
    \hline
\end{tabular}
\vspace{-0.2cm}
\end{table}

\begin{table}[th]
\caption[IRT Resources]{IRT resources}
\begin{tabular}{ll}
  \hline
  \noalign{\smallskip}
Mass telescope (incl. margin)        & 130 kg \\
Mass instrument (optics+electronics) & 150 kg \\
Power electronics                     & 125 W \\
Power LM-band cooling                 & 200 W \\
Raw data / day (2.5x compressed)      & 100 Gbit \\
  \noalign{\smallskip}
  \hline
  \end{tabular}
\vspace{-0.3cm}
\end{table}

The XRM and IRT should be co-aligned such that the IRT FOV is centered
in the FOV of the central X-ray telescope. They should have separate
star-trackers to allow the boresight to be measured/tracked.

The IR detector temperatures need to be stable to within $\pm$0.1K,
thus each requiring a special control loop. The low operating 
temperatures (see critical issue below) and the frequent re-pointings
of the XRM/IRT require special consideration of the thermal architecture.

An optical system with several dichroics in the converging beam (GROND) 
has been built and demonstrated to achieve 0\farcs4 image quality
\cite{gbc08}. Materials for similar systems up to 5$\mu$ are available,
including radiation-hard versions.

\subsection{Nice-to-have additions and option}

There are two additional instruments which have been discussed
in the preparation of this proposal, but are not proposed as 
default instrumentation due to lower priority.

\noindent
(1) {\bf Neutron-Monitor:}
The neutron flux in a LEO varies with time, and induced $\gamma$-radiation
in the telescope is a source of background. A small neutron detector
would act as monitor of the general radiation field, and thus 
substantially help in fighting the background radiation. \\
(2) {\bf Lobster X-ray monitor:} The Lobster camera principle is
ideal for an all-sky monitor at soft X-rays with a spatial
resolution in the few-arcmin range. It would allow detection of
the prompt X-rays which are connected to the GRB prompt emission.
A Lobster system, properly adapted to the GRIPS needs (ring-like
FOV), would allow the localisation of (i) the $\sim$30\% of GRBs
for which the GRM will not be able to measure a position and (ii)
those GRBs for which Earth or Sun constraints would prohibit
slewing the XRT. 

\section{Summary}

GRIPS would be an extraordinary tool to advance the study of the 
{\it nonthermal and violent Universe}.   

The GRIPS mission would provide the data to answer key questions of high-energy 
astrophysics. Moreover, the all-sky survey with an expected number of more 
than 2000 sources, many of them new, will at the same time
serve a diversity of communities for the astronomical exploration of so-far 
unidentified X/$\gamma$-ray sources and of new phenomena. 
The delivery of triggers on bursting sources of high-energy emission will 
amplify the scientific impact of GRIPS across fields and communities.
As the 2010 Decadal Survey Report of the US Academy of Science puts it, 
``Astronomy is still as
much based on discovery as it is on predetermined measurements."


\end{document}